\documentclass[12pt,aps,preprint,prd,nofootinbib,groupedaddress]{revtex4-1}
\usepackage{hyperref}
\usepackage{amssymb}
\usepackage{amsmath}
\usepackage{bm}
\usepackage{graphicx}
\usepackage{mathtools}

\def\be{\begin{equation}}
\def\te{\end{equation}}
\def\ee{\end{equation}}
\def\ba{\begin{eqnarray}}
\def\bea{\begin{eqnarray}}
\def\nn{\nonumber\\}
\def\tea{\end{eqnarray}}
\def\ea{\end{eqnarray}}
\def\eea{\end{eqnarray}}

\begin{document}

%
%

\title{Causal Relativistic Hydrodynamics of Conformal Fermi-Dirac Gases}

\author{Milton Aguilar}
\email[E-mail me at: ]{ mil@df.uba.ar}
\author{Esteban Calzetta}
\email[E-mail me at: ]{calzetta@df.uba.ar}
\affiliation{Universidad de Buenos Aires. Facultad de Ciencias Exactas y Naturales. Departamento de F\'isica. Buenos Aires, Argentina. \\ CONICET - Universidad de Buenos Aires. Facultad de Ciencias Exactas y Naturales. Instituto de F\'isica de Buenos Aires (IFIBA). Buenos Aires, Argentina.}


\begin{abstract}
In this paper we address the derivation of causal relativistic hydrodynamics, formulated within the framework of Divergence Type Theories (DTTs), from kinetic theory for spinless particles obeying Fermi-Dirac statistics. The approach leads to expressions for the particle current and energy momentum tensor that are formally divergent, but may be given meaning through a process of regularization and renormalization. We demonstrate the procedure through an analysis of the stability of an homogeneous anisotropic configuration. In the DTT framework, as in kinetic theory, these configurations are stable. By contrast, hydrodynamics as derived from the Grad approximation would predict that highly anisotropic configurations are unstable.

\end{abstract}
\pacs{25.75.-q, 24.10.Nz, 24.10.-i}
\maketitle



\section{Introduction}
The successful application of relativistic hydrodynamics  \cite{Eck40,LL6,ISR76,IsraelStewart,ISR88,ReZa13,Sch14} to the description of high energy heavy ion collisions \cite{Rom10,Flor10,CalzettaVdL,Monnai14} has led not only to a revival of this theory, but also to the demand of enlarging its domain of applicability to regimes where the system of interest is still away from local thermal equilibrium \cite{Str14b,JeHe15,Rom16}, and so the usual strategy of deriving hydrodynamics as an expansion in deviations from ideal behavior is not available. Moreover, the best known implementations of this strategy, namely the Chapman-Enskog \cite{Chapman,Enskog} and Grad \cite{Grad1,Grad2} approximations, face severe problems, such as spurious instabilities, as it will be shown below. 

The so-called divergence type theories (DTTs)\cite{liu72,LiuMullerRuggeri,GerochLindblom1,GerochLindblom2} are an appealing alternative for the derivation of relativistic causal hydrodynamics because in this framework both the conservation laws for particle number and the energy-momentum tensor as well as the Second Law of Thermodynamics are rigorous properties of the theory, no matter how far from ideal behavior. For this reason the solutions of the theory may be trusted to be at least qualitatively faithful to the underlying kinetic theory. By contrast, a formalism that only enforces the Second Law in an approximate way could lead to unphysical results if the system makes a large excursion away from local thermal equilibrium, even if it is a transient one, and then the whole further evolution would be compromised. Furthermore, in more complete theories including gauge fields\cite{MaMr06,PRD12a,AIP14}, these spurious instabilities could mask or get entangled with legitimate plasma instabilities \cite{Sch06,mama07,PRC13a,MSS16,KC16}.

In this paper we shall analyze the derivation of DTT relativistic hydrodynamics from kinetic theory taking as test case a gas of spinless and massless particles obeying Fermi-Dirac statistics \cite{LewisRomatschke}. In the Grad approach, this derivation consists on formulating an ansatz for the one particle distribution function (1pdf), parameterized by the hydrodynamic variables. Later on the hydrodynamic currents such as the particle number current and the energy-momentum tensor are derived as moments of the parameterized 1pdf, and the corresponding equations as moments of the Boltzmann equation \cite{DKR10,DMNR12a,DMNR12b,Cristian}. This procedure may be replicated in the DTT framework, but it leads to formally divergent expressions. Therefore it is necessary to interpolate a process of regularization and renormalization by which these expressions become meaningful. The conclusion is that a DTT can be derived from kinetic theory, but not uniquely.

As a demonstration of the formalism we shall carry on the procedure, adopting a regularization and renormalization scheme that does not introduce new dimensionful parameters in the theory, preserves positive expressions, and gives the right results in equilibrium, where all relevant expressions are finite to begin with. We shall use the resulting DTT to investigate the stability of an anisotropic (though axisymmetric) homogeneous configuration (a precise characterization will be given below). These are always stable in kinetic theory, but we will show that the Grad approximation predicts an instability  if the anisotropy exceeds a certain threshold. DTT agrees with kinetic theory predicting again stability. As it ought to be expected, the quantitative agreement worsens for larger deviations form equilibrium. 

Let us be more specific about the contents of this paper. We consider a gas of massless, spinless particles obeying Fermi-Dirac statistics. In relativistic kinetic theory \cite{S71,I72,IsrSt79,dGvLW,CalHu08}, the state of the gas is described by a 1pdf $f=f\left(x^{\mu},p_{\nu}\right)$, where the momentum variable is restricted to the positive mass shell $p^2=0=\vec{p}^2-p^{02}$, $p^0\ge 0$ (we adopt the $\left(-+++\right)$ signature for Minkowsky metric  $\eta_{\mu\nu}$). From $f$ we derive the energy-momentum tensor
\be
T^{\mu\nu}=\int\;Dp\;p^{\mu}p^{\nu}f
\label{EMT}
\te
where $Dp$ is the invariant measure
\be 
Dp=\frac{2dp_{0}d^3p}{\left(2\pi\right)^3}\delta\left(p^2\right)\theta\left(p^0\right)
\te
Observe that $T^{\mu\nu}$ is traceless. $T^{\mu\nu}$ admits one (and only one) timelike eigenvector
\be 
T^{\mu\nu}u_{\nu}=-\rho\; u^{\mu}
\te
$u^2=-1$. We say $u^{\mu}$ is the (Landau-Lifshitz) fluid velocity \cite{LL6}, and $\rho$ the energy density. The other relevant current is the entropy flux
\be
S^{\mu}=-\int\;Dp\;p^{\mu}\left\{\left(1-f\right)\ln\left(1-f\right)+f\ln f\right\}
\label{entropy}
\te
and $s=-u_{\mu}S^{\mu}$ is the entropy density. In equilibrium, the 1pdf must maximize the entropy density for a given energy density. This obtains when $f$ is the Fermi-Dirac distribution
\be
f_{eq}=\frac1{e^{-\beta_{\mu}p^{\mu}}+1}
\label{FD}
\te
where $\beta_{\mu}=u_{\mu}/T$, $T$ being the temperature. Thus in equilibrium 
\bea
T_{eq}^{\mu\nu}&=&\sigma_{SB}T^4\left[u^{\mu}u^{\nu}+\frac13\Delta^{\mu\nu}\right]\nn
S_{eq}^{\mu}&=&\frac 43\sigma_{SB}T^3u^{\mu}
\label{eqcur}
\tea
where $\Delta^{\mu\nu}=\eta^{\mu\nu}+u^{\mu}u^{\nu}$ and $\sigma_{SB}=7 \pi^{2} / 240$ is Stefan-Boltzmann's constant.

Out of equilibrium $f$ evolves according to the Boltzmann equation \cite{S71,I72,IsrSt79,dGvLW,CalHu08}
\be
p^{\nu}\frac{\partial}{\partial x^{\nu}}f=I_{col}\left[f\right]
\label{Boltzmann}
\te
The collision integral $I_{col}$ vanishes in equilibrium, and obeys 
\be
\int\;Dp\;p^{\mu}\;I_{col}\left[f\right]=0
\te 
which enforces energy-momentum conservation 
\be
T^{\mu\nu}_{,\nu}=0
\label{conservation}
\te
and the $H$ theorem
\be
\int\;Dp\;\ln\left[\frac 1f-1\right]\;I_{col}\left[f\right]\ge 0
\te 
which enforces the Second Law 
\be
S^{\mu}_{,\mu}\equiv\sigma\ge 0.
\label{secondlaw}
\te
For concreteness we shall adopt the Anderson-Witting collision term\cite{AndersonWitting1,AndersonWitting2,TI10}
\be 
I_{col}\left[f\right]=\frac{1}{\tau} u_{\mu} p^{\mu} \left( f - f_{eq} \right).
	\label{kinetic_equation}
\te
where $f_{eq}$ is the equilibrium distribution with the same velocity and energy density as the nonequilibrium 1pdf $f$. Another frequently used prescription is the Marle or BGK one \cite{Marle1,Marle2}, where $-u_{\mu} p^{\mu}$ in the right hand side is replaced by a power of temperature. 

For a general 1pdf, $T^{\mu\nu}$ acquires a new term, the viscous energy-momentum tensor $\Pi^{\mu\nu}$, 
\be
T^{\mu\nu}=\sigma_{SB}T^4\left[u^{\mu}u^{\nu}+\frac13\Delta^{\mu\nu}\right]+\Pi^{\mu\nu}
\te
Since $\Pi^{\mu\nu}$ is traceless and transverse $\Pi^{\mu}_{\mu}=u_{\nu}\Pi^{\mu\nu}=0$ it has $5$ independent components, elevating the total number of degrees of freedom in $T^{\mu\nu}$ to $9$. The four conservation equations Eq. (\ref{conservation}) are therefore not enough to predict the evolution of the energy-momentum tensor. The problem of relativistic hydrodynamics is to provide the missing equations.

The Chapman-Enskog approach\cite{Chapman,Enskog} assumes that at every point $f$ is close to an equilibrium distribution, although with position dependent temperature and velocity. Then a solution of Eq. (\ref{Boltzmann}) is sought as a formal expansion in powers of the relaxation time $\tau$ introduced in Eq. (\ref{kinetic_equation})
\be
f_{Ch-E}=f_{eq}\left[1+{\tau}\left(1-f_{eq}\right)\delta f_{Ch-E}\right]
\label{Ch-E}
\te
Inserting this into Eq. (\ref{Boltzmann}) with collision term (\ref{kinetic_equation}), and using the conservation equations Eq. (\ref{conservation}) to order $\tau^0$ to simplify the result, we obtain to lowest order 
\be
\delta f_{Ch-E}=\frac{-1}{2T\left|u_{\rho}p^{\rho}\right|}\sigma_{\mu\nu}p^{\mu}p^{\nu}
\te
where we introduced the shear tensor
\be 
\sigma_{\mu\nu}=\Delta^{\rho}_{\mu}\Delta^{\lambda}_{\nu}\left[u_{\rho,\lambda}+u_{\lambda,\rho}-\frac23\Delta_{\rho\lambda}u^{\tau}_{,\tau}\right]
\te
A straightforward computation yields
\be
\Pi^{\mu\nu}=-\eta\sigma^{\mu\nu}
\te 
where $\eta= \left( 7 \pi^{2} / 900 \right) \tau T^4$ is the shear viscosity. Thus in this approach the viscous energy-momentum tensor is slaved to the degrees of freedom that describe the ideal fluid at $\tau =0$. This eventually leads to a parabolic system of equations of motion, incompatible with relativistic causality \cite{HisLin83,HiscockLindblom,HisLin88}.

To overcome this difficulty, the Grad approach\cite{Grad1,Grad2} proposes instead a 1pdf
\begin{equation}
	f = f_{eq} \left[ 1 +  Z\right]
	\label{Grad_f}
\end{equation}
\be
Z=\frac{1}{2 T \left( - u_{\mu} p^{\mu} \right)} \,\left( 1 - f_{eq} \right) \xi_{\mu \nu} p^{\mu} p^{\nu} 
\label{Grad_Z}
\te
$\xi_{\mu\nu}$ is traceless and transverse, and it is regarded as an independent tensorial degree of freedom. It is directly related to $\Pi_{\mu\nu}$, since 
\be 
\Pi_{\mu\nu}= \left(\frac{7 \pi^{2}} { 900 }\right) T^4 \xi_{\mu\nu}
\label{Gradpi}
\te 
Since $\xi_{\mu\nu}$ is not positive definite, the Grad approximation will lead to negative pressures if $\xi_{\mu\nu}$  is large enough, which underscores the unapplicability of the theory far from equilibrium. Moreover, we shall show below the theory has spurious instabilities even before that limit is reached.

To obtain a dynamics for these new $5$ degrees of freedom in the viscous energy-momentum tensor, one further moment of the Boltzmann equation is computed. The first moments yield energy-momentum conservation Eq. (\ref{conservation}). Instead of the ten second moments, we only consider the traceless, transverse ones
\be
\left[\Delta^{\rho}_{\mu}\Delta^{\lambda}_{\nu}-\frac13\Delta^{\rho\lambda}\Delta_{\mu\nu}\right]\left\{A^{\sigma\mu\nu}_{,\sigma}-I^{\mu\nu}\right\}=0
\label{NEC}
\te
where
\bea 
A^{\sigma\mu\nu}&=&\int\;Dp\;p^{\mu}p^{\nu}p^{\sigma}f\nn
I^{\mu\nu}&=&\int\;Dp\;p^{\mu}p^{\nu}I_{col}
\label{dtteq}
\tea
The nonequilibrium current $A^{\sigma\mu\nu}$ is totally symmetric and traceless on any two indexes. This approach leads to a Maxwell-Cattaneo \cite{HeatWaves} equation for $\xi^{\mu\nu}$ and enforces causality. However, it cannot be applied arbitrarily far from equilibrium, because, as we shall show below, it predicts instabilities that do not exist in the kinetic theory. For further discussion of the Chapman-Enskog and Grad approaches see \cite{Jais13}

The DTT framework keeps Eqs. (\ref{conservation}) and (\ref{NEC}) as the fundamental equations, but now seeks a 1pdf which maximizes entropy density for given energy density and $A^{0ij}$ components in the rest frame. This leads to the introduction of a new tensor Lagrange multiplier $\zeta_{\mu\nu}$ besides $T$ and $u_{\mu}$. Assuming  $\zeta_{\mu\nu}$ is symmetric, traceless and transverse, it is equivalent to $5$ new degrees of freedom. Thus the theory has the same number of degrees of freedom as the energy-momentum tensor, with the nonequilibrium current slaved to it (we will return to this point below). This means that in the DTT, two evolutions starting with the same energy-momentum tensor will remain identical, though it is known that they may diverge in kinetic theory \cite{Ollitrault}. Even so, we will show that the DTT outperforms the Grad approximation, in the sense that it is free from the spurious instabilities that appear in the latter.

The problem is that, though the variational problem leading to the DTT 1pdf is easily solved, the formal expressions one obtains for the energy-momentum tensor and the nonequilibrium current diverge \cite{ReulaNagy}. Thus it is necessary to regularize and renormalize them to make sense of the theory. This adds a new, non unique stage in the derivation of hydrodynamics from kinetic theory. Our goal is to show a concrete procedure to obtain finite quantities for the relevant currents, and then to use this procedure to demonstrate the stability of anisotropic, axisymmetric configurations, in agreement with kinetic theory.

The rest of the paper is organized as follows. In next section we provide some further background on the DTT framework, starting from the purely macroscopic point of view whereby it was first introduced, and then linking it to kinetic theory. Then we proceed to regularize and renormalize the formal expressions for the energy-momentum tensor and nonequilibrium current. Section 3 provides a first comparison of DTT and Grad hydrodynamics through the analysis of the pressure anisotropy; we show that while in Grad hydrodynamics the pressure anisotropy becomes negative when far from equilibrium, in DTT it is bounded below. Section 4 is the main part of this paper; here we discuss the stability of anisotropic homogeneous configurations, comparing the analysis made within DTT and Grad hydrodynamics to the one in kinetic theory. We conclude that while both kinetic theory and DTT predict anisotropic axisymmetric configurations are always stable, Grad hydrodynamics shows an instability if the anisotropy is large enough. We conclude with some brief final remarks. 

The two appendices discuss important conceptual issues. Appendix A presents a general framework to analyze stability of conformal hydrodynamic theories of the type discussed in this paper. We show that the instability of the Grad approximation is a consequence of the linearization of the one-particle distribution function with respect to $\xi^{\mu\nu}$. In Appendix B we show how the regularization and renormalization scheme presented here may be applied to fluids obeying Maxwell-J\"uttner or Bose-Einstein statistics.

\section{Divergence Type Theories}
DTTs are theories in which all the dynamical equations can be written as divergences of tensor fields. They were originally developed by Liu, M\"uller and Ruggeri\cite{liu72,LiuMullerRuggeri} as a response to the perceived flaws of the so-called ``first order'' relativistic hydrodynamics of Eckart \cite{Eck40} and Landau-Lifshitz \cite{LL6}. They were later extended by the works of Geroch and Lindblom\cite{GerochLindblom1,GerochLindblom2} and Reula and Nagy\cite{ReulaNagy}. They were applied to study relativistic hydrodynamic fluctuations in \cite{Calzetta_CQG}, and free streaming flows in \cite{PRD15}. They were applied to the study of relativistic heavy ion collisions in \cite{PRC09,PRC10b,IJMPD10}.

The main fields on a DTT are the particle-number current $N^{\mu}$ and the energy-momentum tensor $T^{\mu \nu}$ and their dynamics are governed by the conservation equations
\begin{equation}
	\begin{dcases}
		\partial_{\mu} N^{\mu} = 0 \\
		\partial_{\mu} T^{\mu \nu} = 0
	\end{dcases}
	\label{DDT_dynamics1}
\end{equation}
Closure of the system is achieved by the addition of the balance law of fluxes
\begin{equation}
	\partial_{\mu} A^{\mu \nu \rho} = I^{\nu \rho},
	\label{DDT_dynamics2}
\end{equation}
\noindent where $A^{\mu \nu \rho}$ and $I^{\nu \rho}$ are algebraic functions of $N^{\mu}$ and $T^{\mu \nu}$. This means we are not adding extra degrees of freedom.
To relate these currents among themselves, it is assumed not only that there is an entropy flux vector $S^{\mu}$ (cfr. Eq. (\ref{entropy})) whose divergence $\sigma$ (cfr. Eq. (\ref{secondlaw})) is positive, but moreover that both $S^{\mu}$ and  $\sigma$ are algebraic functions of $N^{\mu}$ and $T^{\mu \nu}$, such that the positivity of $\sigma$ follows from Eqs. \eqref{DDT_dynamics1} and \eqref{DDT_dynamics2} alone. It can be shown \cite{liu72} that this implies the existence of  a vector $\chi^{\mu} = \chi^{\mu} \left( \alpha , \beta_{\mu} , \zeta_{\mu \nu} \right)$ and a source $I^{\mu \nu} = I^{\mu \nu} \left( \alpha , \beta_{\mu} , \zeta_{\mu \nu} \right)$ in such a way that the fields $N^{\mu}$, $T^{\mu \nu}$ and $A^{\mu \nu \rho}$ can be computed as the following partial derivatives
\begin{equation}
	N^{\mu} = \frac{\partial \chi^{\mu}}{\partial \alpha}, \;\;\; T^{\mu \nu} = \frac{\partial \chi^{\mu}}{\partial \beta_{\nu}} \;\;\; \text{and} \;\;\; A^{\mu \nu \rho} = \frac{\partial \chi^{\mu}}{\partial \zeta_{\nu \rho}}.
\end{equation}
\noindent The variables $\alpha$ and $\beta_{\mu}$ are related to the chemical potential, the hydrodynamic velocity and the temperature and the symmetric tensor $\zeta_{\mu \nu}$ provides the necessary degrees of freedom to match any given energy-momentum tensor. 
\subsection{Dissipative Type Theories from Kinetic Theory}
Since a DTT is totally defined by the generating function $\chi^{\mu}$, to establish a link with kinetic theory it is necessary to relate the generating function to the 1pdf. Let us consider the massless case from now on, so we shall drop $\alpha$ and the particle number current from the discussion.

In equilibrium, the energy-momentum tensor Eq. (\ref{eqcur}) may be recovered from the generating function
\begin{equation}
	\chi^{\mu}_{eq} = - \int Dp \, p^{\mu} \, \text{ln} \left( 1 - f_{eq} \right),
\end{equation}
where $f_{eq}$ is the Fermi-Dirac 1pdf Eq. (\ref{FD}). This suggests to generalize this to the dissipative case by writing 
\begin{equation}
	\chi^{\mu} = - \int Dp \, p^{\mu} \, \text{ln} \left( 1 - f \right)
	\label{generating}
\end{equation}
\noindent and a source chosen to match the Anderson-Witting collision term, as in Eq. (\ref{dtteq}). In order that we may recover the nonequilibrium current as a derivative of $\chi^{\mu}$ we must write $f$ as a deformation of the Fermi-Dirac distribution \cite{ReulaNagy}
\begin{equation}
	f \left[ p^{\mu} , u^{\mu} , \zeta^{\mu \nu} , T \right] = \frac{1}{e^{- \frac{1}{T} u_{\mu} p^{\mu} - \zeta_{\mu \nu} p^{\mu} p^{\nu}} + 1}
	\label{FDD}
\end{equation}
This distribution function maximizes the entropy density for given values of $T^{00}$ and $A^{0ij}-\left(1/3\right)\delta^{ij}A^{0k}_k$, as measured in the fluid rest frame. $f_{eq}$ in the collision term Eq. (\ref{kinetic_equation}) reads
\begin{equation}
	f_{eq} \left[ p^{\mu} , u^{\mu} , T_{eq} \right] = \frac{1}{e^{- \frac{1}{T_{eq}} u_{\mu} p^{\mu}} + 1}.
	\label{FDeq}
\end{equation}
\noindent with $T$, $\zeta^{\mu \nu}$ and $T_{eq}$ related by
\begin{equation}
	\rho \left( T,\zeta^{\mu \nu} \right) =\sigma_{SB} T_{eq}^4.
	\label{relationT}
\end{equation}
\indent Although the theory can be formally defined as is and the energy-momentum tensor $T^{\mu \nu}$ and nonequilibrium current $A^{\mu \nu \rho}$ can be expressed as partial derivatives of the generating function $\chi^{\mu}$, there's an obvious problem in Eq. \eqref{generating}. Since the quadratic form $\zeta_{\mu \nu} p^{\mu} p^{\nu}$ is not negative definite, there are values of $\zeta_{\mu \nu}$ such that
\begin{equation}
	f \xrightarrow[p^{\mu} \rightarrow \infty] \;\;\; 1,
	\label{sing1}
\end{equation}
\noindent making the generating function divergent. Since these are all states with occupation number one, this singularity can be interpreted as the Dirac Sea. There's also a not-so-obvious singularity because of the behavior of the integrals near the manifold where the quadratic part of the argument of the exponential becomes zero. That is, the manifold defined by the equation
\begin{equation}
	\zeta_{\mu \nu} p^{\mu} p^{\nu} = 0.
	\label{sing2}
\end{equation}
This will be further clarified below.
\subsection{Regularization}
\begin{figure}[h]
	\centerline{\includegraphics[width=17cm]{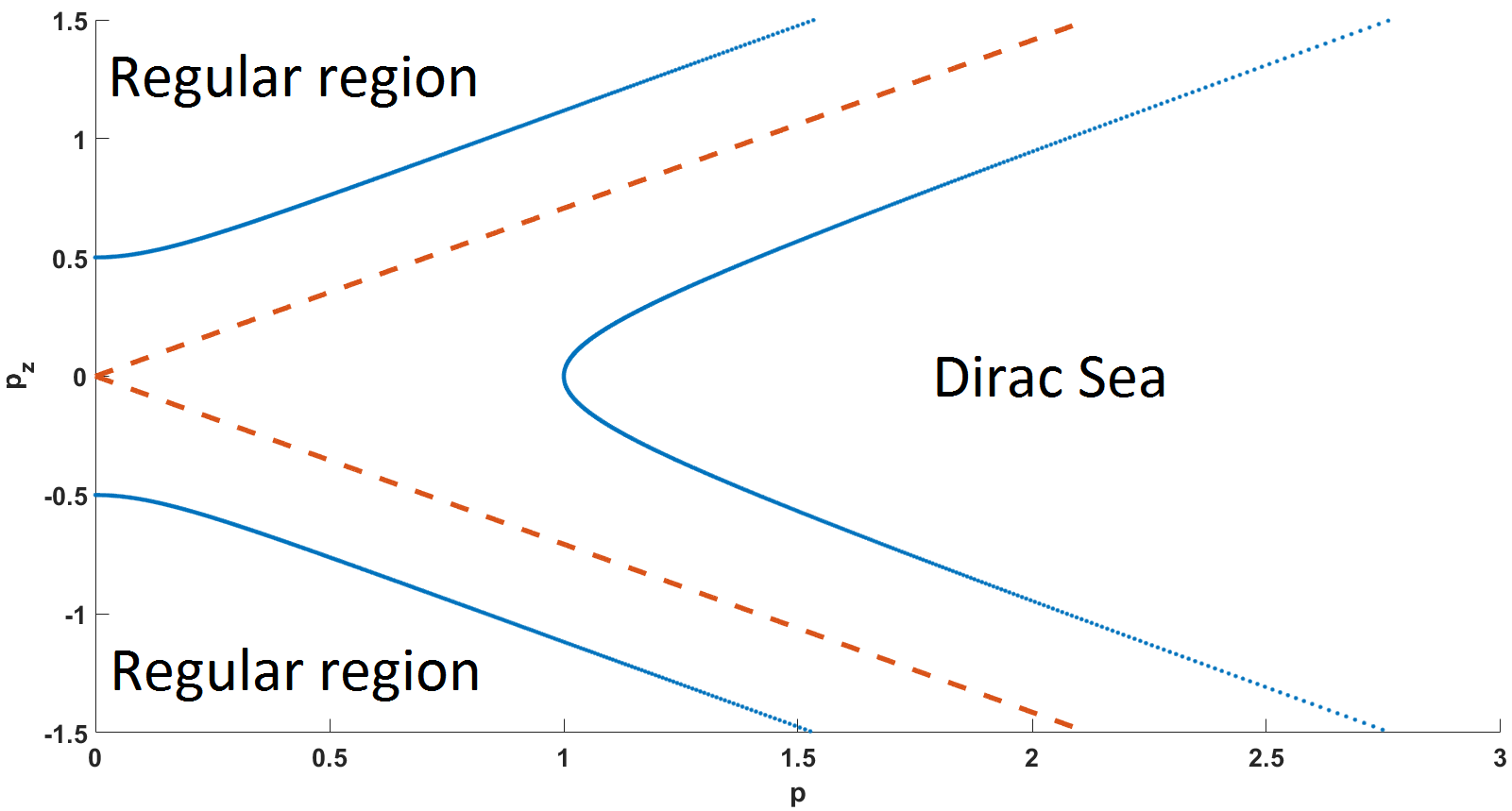}}
		\caption{A graphic of the $p p_{z}$-plane showing the singularities at $\theta_{0}$ in dashed lines as well as a transverse cut of the surface befined by the equation $p - p^{2} \left( 1 - 3 \text{cos}^{2} \theta \right) = 0$ as an example. \label{diracSea}}
\end{figure}

In order to take care of the singularities Eq. \eqref{sing1} and Eq. \eqref{sing2}, let's fix the dissipative tensor $\zeta_{\mu \nu}$. Motivated by the family of solutions first introduced by Romatschke and Strickland \cite{RomatschkeStrickland} (see also \cite{Florkowski,Tinti}), we choose the transverse, traceless and axisymmetric case
\begin{equation}
	T^{2} \zeta_{\mu \nu} = \text{diag} \left( 0 , \zeta_{0} , \zeta_{0} , - 2 \zeta_{0} \right)
\end{equation}
\noindent($\zeta_0>0$) so the type of integrals to regularize are
\begin{equation}
	I_{1} \left[ g \right] = \int_{0}^{\infty} dp \int_{0}^{\pi / 2} d \theta \, g(p,\theta) \, f \left( p , \theta \right)
	\label{Regularizacion_I1}
\end{equation}
\noindent and
\begin{equation}
	I_{2} \left[ g \right] = \int_{0}^{\infty} dp \int_{0}^{\pi / 2} d \theta \, g(p,\theta) \, f \left( p , \theta \right) \, \left[ 1 - f \left( p , \theta \right) \right],
	\label{Regularizacion_I2}
\end{equation}
\noindent where $g$ is a polynomial function in the variables $p$, $\text{cos} \, \theta$ and $\text{sin} \, \theta$ and $f$ is the dissipative Fermi-Dirac distribution Eq. \eqref{FDD},
\begin{equation}
	f \left( p , \theta \right) = \frac{1}{ e^{p - \zeta_{0} p^{2} \left( 1 - 3 \text{cos}^{2} \theta \right)} + 1}.
\end{equation}
Note that the first singularity, Eq. \eqref{sing1}, happens when $\text{cos}^2 \theta < 1/3$ and the second one, Eq. \eqref{sing2}, when $\text{cos}^2 \theta \rightarrow 1/3^{-}$ so the $p p_{z}$-plane gets divided into two sections delimited by $\text{cos}^2 \theta_{0} = 1/3$. On one of those sections, when $\text{cos}^2 \theta > 1/3$, the integral is regular and on the other one is where the singularities are located. Figure (\ref{diracSea}) shows this.
The Dirac Sea singularity, namely Eq. \eqref{sing1}, can be eliminated by a simple integration by parts. The surface term, which is infinite, is discarded and the $f \left( p , \theta \right)$ that used to be in the integrand gets replaced by $f \left( p , \theta \right) \, \left[ 1 - f \left( p , \theta \right) \right]$ in the remaining term, which goes to zero as $f$ goes to one. The other singularity, Eq. \eqref{sing2}, requieres further analysis. The goal of this section is to introduce a regularization procedure which eliminates both singularities but without introducing new parameters to the theory, as to preserve the conformal invariance.
\subsubsection{The $I_{1}$ Case}
Let's start by dividing the  $I_{1}$ integral in $\theta_{0} = \text{cos}^{-1} \left( 1 / \sqrt{3} \right)$,
\begin{gather}
	I_{1} \left[ g \right] = I_{1}^{<} \left[ g \right] + I_{1}^{>} \left[ g \right] \notag \\
	\doteq \int_{0}^{\infty} dp \int_{0}^{\theta_{0}} d \theta \, g(p,\theta) \, f \left( p , \theta \right) + \int_{0}^{\infty} dp \int_{\theta_{0}}^{\pi / 2} d \theta \, g(p,\theta) \, f \left( p , \theta \right).
\end{gather}
$I_{1}^{<}$ is finite and is left as is. To study the $I_{1}^{>}$ term, define a function $G$ by
\begin{equation}
	G \left( p , \theta \right) \doteq \int_{\theta}^{\pi / 2} d \phi \, g ( p , \phi ),
	\label{gfunction}
\end{equation}
\noindent so that $G \left( p , \pi / 2 \right) = 0$ and
\begin{equation}
	\frac{\partial G}{\partial \theta} = - g( p , \theta ).
\end{equation}
Now integrate by parts to obtain
\begin{gather}
	I_{1}^{>} = \int_{0}^{\infty} dp \, \frac{G \left( p , \theta_{0} \right)}{e^{p} + 1}  \notag \\ %
	+ \frac{3}{2} \zeta_{0} \int_{0}^{\infty} dp \int_{\theta_{0}}^{\pi / 2} d \theta \, p^{2} \, G \left( p , \theta \right) \, \frac{\text{cos} \theta \, \text{sin} \theta}{\text{cosh}^{2} \left[ p/2 - \zeta_{0} p^{2} \left( 1 - 3 \text{cos}^{2} \theta \right) / 2 \right]}.
\end{gather}
\noindent it is evident that the first term is finite. Let's call the second term $K_{Z}$. By using the sum of arguments relation of the hyperbolic cosine and realizing it could be written as a partial derivative, it is possible to rewrite $K_{Z}$ as
\begin{gather}
	K_{Z} = \int_{0}^{\infty} dp \int_{\theta_{0}}^{\pi / 2} d \theta \, \frac{G \left( p , \theta \right)}{\text{sinh} \left( p \right)} \notag \\
	\times \frac{\partial}{\partial \theta} \left\{ \frac{1}{ 1 - \text{tanh} \left( p/2 \right) \text{tanh} \left[ \zeta_{0} p^{2} \left( 1 - 3 \text{cos}^{2} \theta \right) / 2 \right] } \right\}.
\end{gather}
\noindent Performing another integration by parts and defining the auxiliary function
\begin{equation}
	\mathcal{K} \left( \zeta_{0} \right) \doteq \int_{0}^{\infty} dp \int_{\theta_{0}}^{\pi / 2} d \theta \, \frac{g \left( p , \theta \right)}{\text{sinh} \left( p \right)} \,  \frac{1}{ 1 - \text{tanh} \left( p/2 \right) \text{tanh} \left[ \zeta_{0} p^{2} \left( 1 - 3 \text{cos}^{2} \theta \right) / 2 \right] },
\end{equation}
\noindent we arrive at the final expression for $K_{Z}$,
\begin{equation}
	K_{Z} = \mathcal{K} \left( 0 \right) \left[ \frac{\mathcal{K} \left( \zeta_{0} \right)}{\mathcal{K} \left( 0 \right)} - 1 \right].
	\label{KZ}
\end{equation}
\noindent The key here is to identify the ratio in Eq. \eqref{KZ} as the mean value
\begin{equation}
	U \left( \zeta_{0} \right) \doteq \frac{\mathcal{K} \left( \zeta_{0} \right)}{\mathcal{K} \left( 0 \right)} = \left\langle \frac{1}{1 - u} \right\rangle = \int_{0}^{1} du \, \frac{F_{1} \left( u \right)}{1 - u},
\end{equation}
\noindent where $F_{1}$ is a probability density function defined as
\begin{gather}
	F_{1} \left( u \right) \doteq \frac{1}{\mathcal{K} \left( 0 \right)} \int_{0}^{\infty} dp \int_{\theta_{0}}^{\pi / 2} d \theta \, \frac{g \left( p , \theta \right)}{\text{sinh} \left( p \right)} \notag \\ 
	\times \delta \left[ \text{tanh} \left( p/2 \right) \text{tanh} \left[ \zeta_{0} p^{2} \left( 1 - 3 \text{cos}^{2} \theta \right) / 2 \right] - u \right]
\end{gather}
\noindent if $ u \in \left[0 , 1 \right)$ and $F_{1} \left( u \right) = 0$ otherwise. Although $F_{1}(1^{+}) = 0$, $F_{1}(1^{-})$ goes to infinity logarithmically and, therefore, $U$ is divergent. The first step to improve this behavior is to replace $U$ by its Cauchy principal value,
\begin{equation}
	U \left( \zeta_{0} \right) \rightarrow 	U_{PV} \left( \zeta_{0} \right) = \text{Re} \left[ \text{PV} \int_{0}^{1} du \, \frac{F_{1} \left( u \right)}{1 - u} \right],
\end{equation}
\noindent which, as a consequence of the Sokhotski-Plemelj-Fox theorem \cite{SPFT}, equals
\begin{equation}
	U_{PV} \left( \zeta_{0} \right) = \text{Re} \left[ \lim_{\varepsilon \to 0^{+}} \int_{0}^{1} du \, \frac{F_{1} \left( u \right)}{1 - u - i \varepsilon} \right].
\end{equation}
\noindent Now let $e^{W \left( t \right)}$ be the characteristic function of $F_{1}$,
\begin{equation}
	e^{W \left( t \right)} \doteq \int_{0}^{1} du \, F_{1} \left( u \right) \, e^{i t u}.
\end{equation}
\noindent Then $U_{PV}$ can be written as
\begin{equation}
	U_{PV} \left( \zeta_{0} \right) = \text{Re} \left[ \lim_{\varepsilon \to 0^{+}} i \,  \int_{0}^{\infty} dt \, e^{ \left[ W \left( t \right) - i t \right]} \, e^{- \varepsilon t} \right],
\end{equation}
\noindent Since $e^{W \left( t \right)}$ is the characteristic function of $F_{1}$, then $W$ is the cumulant-generating function. This means that $W$ has the formal power series expansion
\begin{equation}
	W \left( t \right) = \sum_{n = 1}^{\infty} \frac{\kappa_{n}}{n!} \left( it \right)^{n},
\end{equation}
\noindent where $\kappa_{n}$ is the $n$th cumulant. That is, $\kappa_{1} = \left\langle u \right\rangle$, $\kappa_{2} = \sigma^{2} = \left\langle u^{2} \right\rangle - \left\langle u \right\rangle^{2}$ and so on. 

\subsubsection{The $I_{2}$ Case}
This case is fairly similar to the first one the only difference being, due to possible divergent terms introduced by $G$, one less integration by parts is performed. We start by dividing the integral the same way as before,
\begin{gather}
	I_{2} \left[ g \right] = I_{2}^{<} \left[ g \right] + I_{2}^{>} \left[ g \right] \doteq \int_{0}^{\infty} dp \int_{0}^{\theta_{0}} d \theta \, g(p,\theta) \, f \left( p , \theta \right) \, \left[ 1 - f \left( p , \theta \right) \right] \notag \\ %
	+ \int_{0}^{\infty} dp \int_{\theta_{0}}^{\pi / 2} d \theta \, g(p,\theta) \, f \left( p , \theta \right) \, \left[ 1 - f \left( p , \theta \right) \right].
\end{gather}
\noindent $I_{2}^{<}$ is finite so we focus our attention on $I_{2}^{>}$. Using the sum of arguments relation of the hyperbolic cosine we write
\begin{gather}
	I_{2}^{>} \left[ g \right] = \frac{1}{4} \int_{0}^{\infty} dp \int_{\theta_{0}}^{\pi / 2} d \theta \, g \left( p , \theta \right) \, \frac{1}{\text{cosh}^{2} \left( p/2 \right) \text{cosh}^{2} \left[ \zeta_{0} p^{2} \left( 1 - 3 \text{cos}^{2} \theta \right) / 2 \right]} \notag \\ %
	\times \frac{1}{\left\{ 1 - \text{tanh} \left( p/2 \right) \text{tanh} \left[ \zeta_{0} p^{2} \left( 1 - 3 \text{cos}^{2} \theta \right) / 2 \right] \right\}^{2}}.
\end{gather}
\noindent Just as before we define the auxiliary function
\begin{equation}
	\mathcal{K} \left( \zeta_{0} \right) \doteq \frac{1}{4} \int_{0}^{\infty} dp \int_{\theta_{0}}^{\pi / 2} d \theta \, g \left( p , \theta \right) \, \frac{1}{\text{cosh}^{2} \left( p/2 \right) \text{cosh}^{2} \left[ \zeta_{0} p^{2} \left( 1 - 3 \text{cos}^{2} \theta \right) / 2 \right]}
\end{equation}
\noindent and the probability density function
\begin{gather}
	F_{2} \left( u \right) \doteq \frac{1}{4 \, \mathcal{K} \left( \zeta_{0} \right)} \int_{0}^{\infty} dp \int_{\theta_{0}}^{\pi / 2} d \theta \, g \left( p , \theta \right) \, \frac{1}{\text{cosh}^{2} \left( p/2 \right) \text{cosh}^{2} \left[ \zeta_{0} p^{2} \left( 1 - 3 \text{cos}^{2} \theta \right) / 2 \right]} \notag \\ %
	\times \delta \left[ \text{tanh} \left( p/2 \right) \text{tanh} \left[ \zeta_{0} p^{2} \left( 1 - 3 \text{cos}^{2} \theta \right) / 2 \right] - u \right]
\end{gather}
\noindent if $ u \in \left[0 , 1 \right)$ and $F_{2} \left( u \right) = 0$ otherwise. Now $I_{2}^{>}$ can be written as the following mean value
\begin{equation}
	I_{2}^{>} \left[ g \right] = \mathcal{K} \left( \zeta_{0} \right) \left\langle \frac{1}{\left( 1 - u \right)^{2}} \right\rangle = \mathcal{K} \left( \zeta_{0} \right) \int_{0}^{1} du \, \frac{F_{2} \left( u \right)}{\left( 1 - u \right)^{2}}.
\end{equation}
\noindent $\mathcal{K}$ is obviously finite but
\begin{equation}
	U \left( \zeta_{0} \right) \doteq  \int_{0}^{1} du \, \frac{F_{2} \left( u \right)}{\left( 1 - u \right)^{2}}
\end{equation}
\noindent is not. In order to improve this behavior we replace $U$ by its Cauchy principal value,
\begin{equation}
	U \left( \zeta_{0} \right) \rightarrow 	U_{PV} \left( \zeta_{0} \right) = \text{Re} \left[ \text{PV} \int_{0}^{1} du \, \frac{F_{2} \left( u \right)}{\left( 1 - u \right)^{2}} \right],
\end{equation}
\noindent which could be written as
\begin{equation}
	U_{PV} \left( \zeta_{0} \right) = - \text{Re} \left[ \lim_{\varepsilon \to 0^{+}} \int_{0}^{\infty} dt \, t \, e^{\left[ W \left( t \right) - i t \right]} \, e^{- \varepsilon t} \right],
\end{equation}
\noindent where $W$ is the cumulant-generating function of $F_{2}$.

\subsection{Renormalization}
So far we have been able to rewrite our integrals in a way that singles out the divergent factors. We must now renormalize them in such a way as to obtain finite expressions.

The idea is to generate a series of expressions for $U_{PV}$ by replacing $F_{1}$ by another distribution function, better behaved than $F_{1}$ as $u\to 1$, but whose irreducible moments agree with those of $F_{1}$ up to a certain order. It is important that we replace $F_{1}$ by another positive function, since this preserves positivity, and it is important that the replacement pdf includes no dimensionful parameters not present in $F_{1}$, since otherwise conformal invariance would be spoiled. The simplest such replacement would be a $\delta$ function with support at $\left\langle u \right\rangle$. In this paper, we shall restrict ourselves to the next approximation, where both $\left\langle u \right\rangle$ and $\sigma^2$ are retained, and $F_{1}$ is replaced by a Gaussian pdf. This is equivalent to considering only the leading term in a Gram - Charlier approximation to $F_1$ \cite{Cramer57}; it must be recalled that Gram - Charlier series, when truncated at higher orders, may not be a true distribution function because it may not be nonnegative definite.  We shall discuss the accuracy of this lowest order approximation below. 

\begin{figure}[h]
	\centerline{\includegraphics[width=17cm]{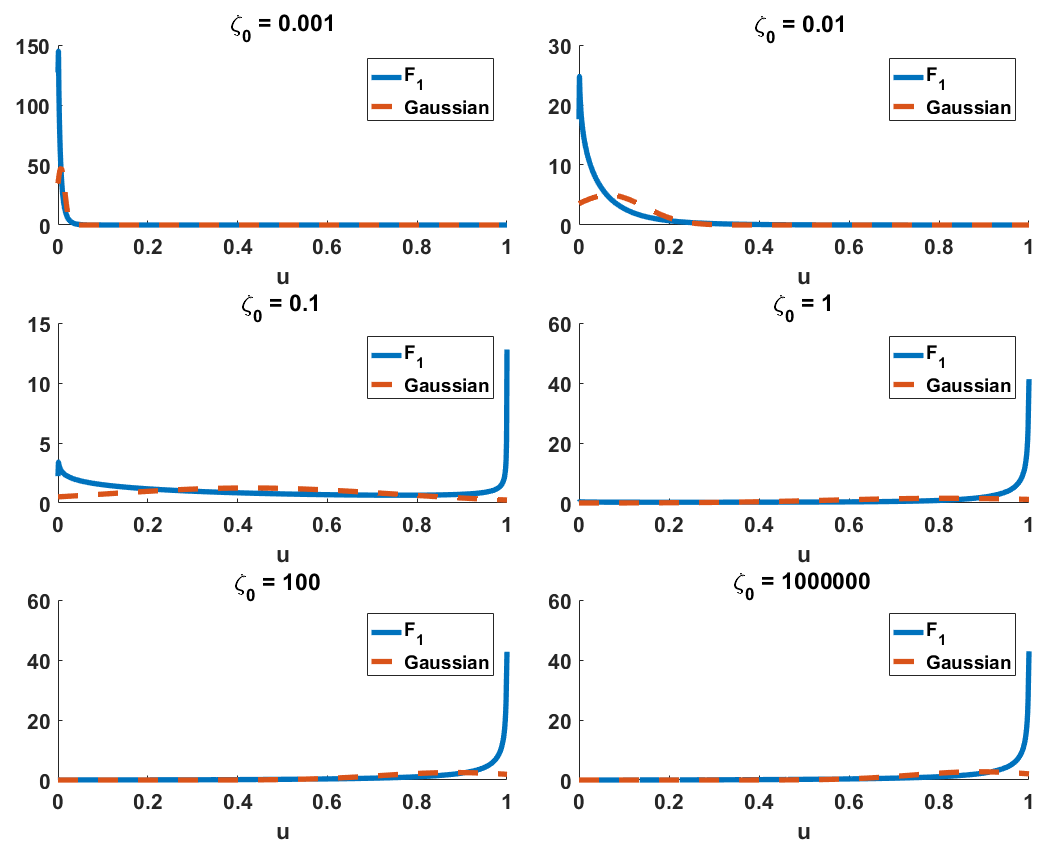}}
	\caption{Comparison between $F_{1}$ (full line) and its Gaussian approximation $\mathcal{F}_{1}$ (dashed line) for different $\zeta_{0}$ values. \label{F1}}
\end{figure}

\begin{figure}[h]
	\centerline{\includegraphics[width=17cm]{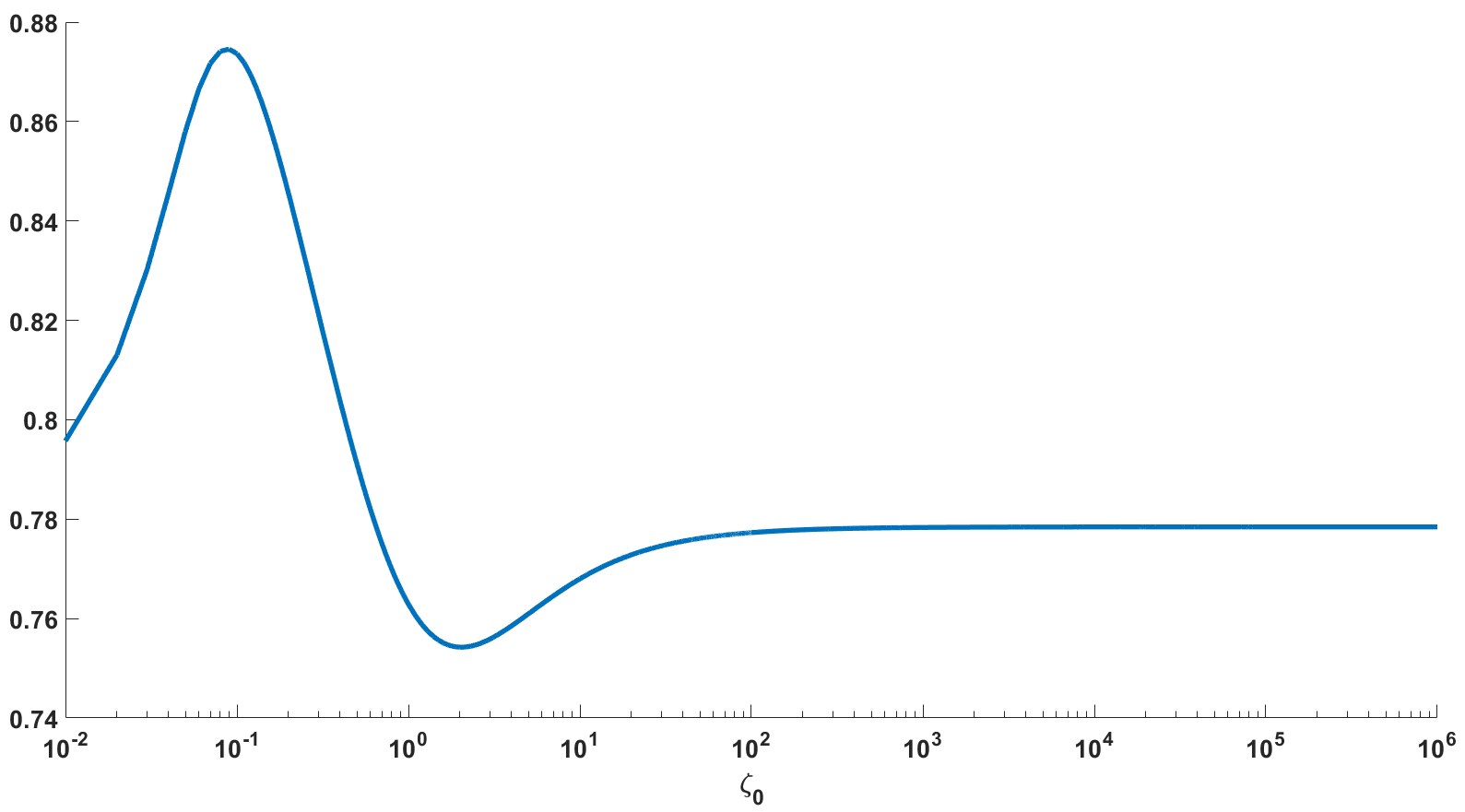}}
	\caption{Integral of $\mathcal{F}_{1}$ over $\left[ 0 , 1 \right]$. $\zeta_{0}$ axis in logarithmic scale. \label{F1_integral}}
\end{figure}

\begin{figure}[h]
	\centerline{\includegraphics[width=17cm]{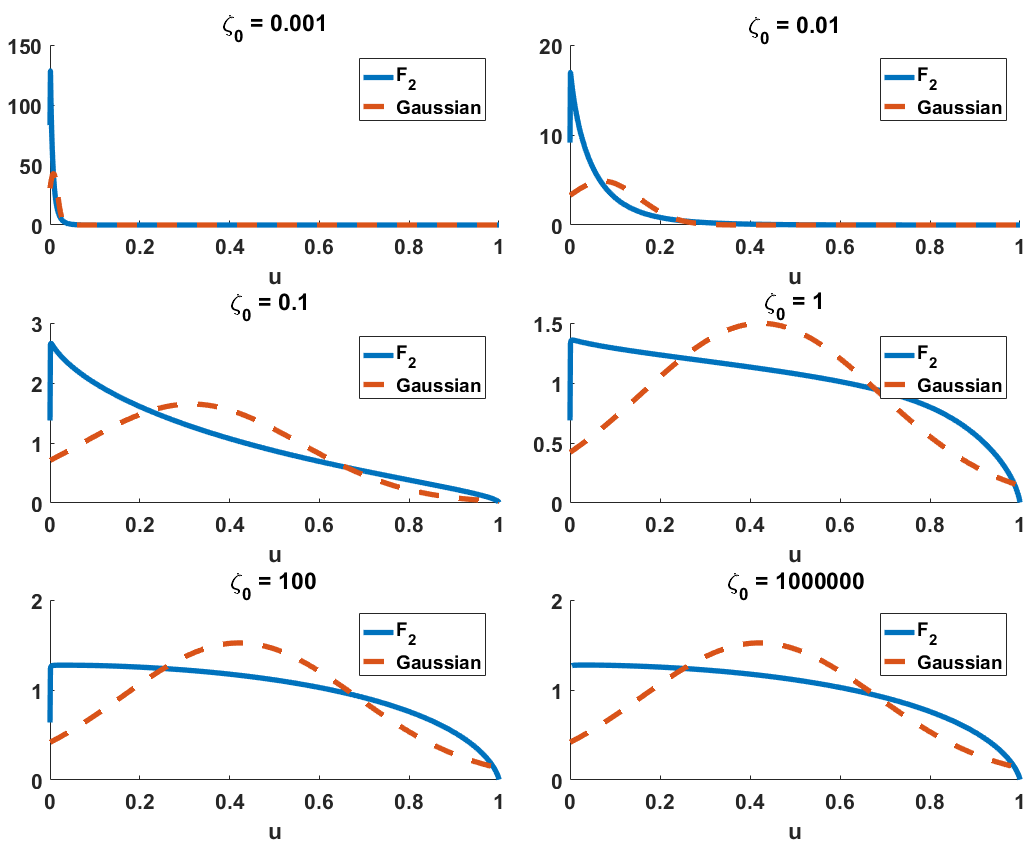}}
	\caption{Comparison between $F_{2}$ (full line) and its Gaussian approximation $\mathcal{F}_{2}$ (dashed line) for different $\zeta_{0}$ values. \label{F2}}
\end{figure}

\begin{figure}[h]
	\centerline{\includegraphics[width=17cm]{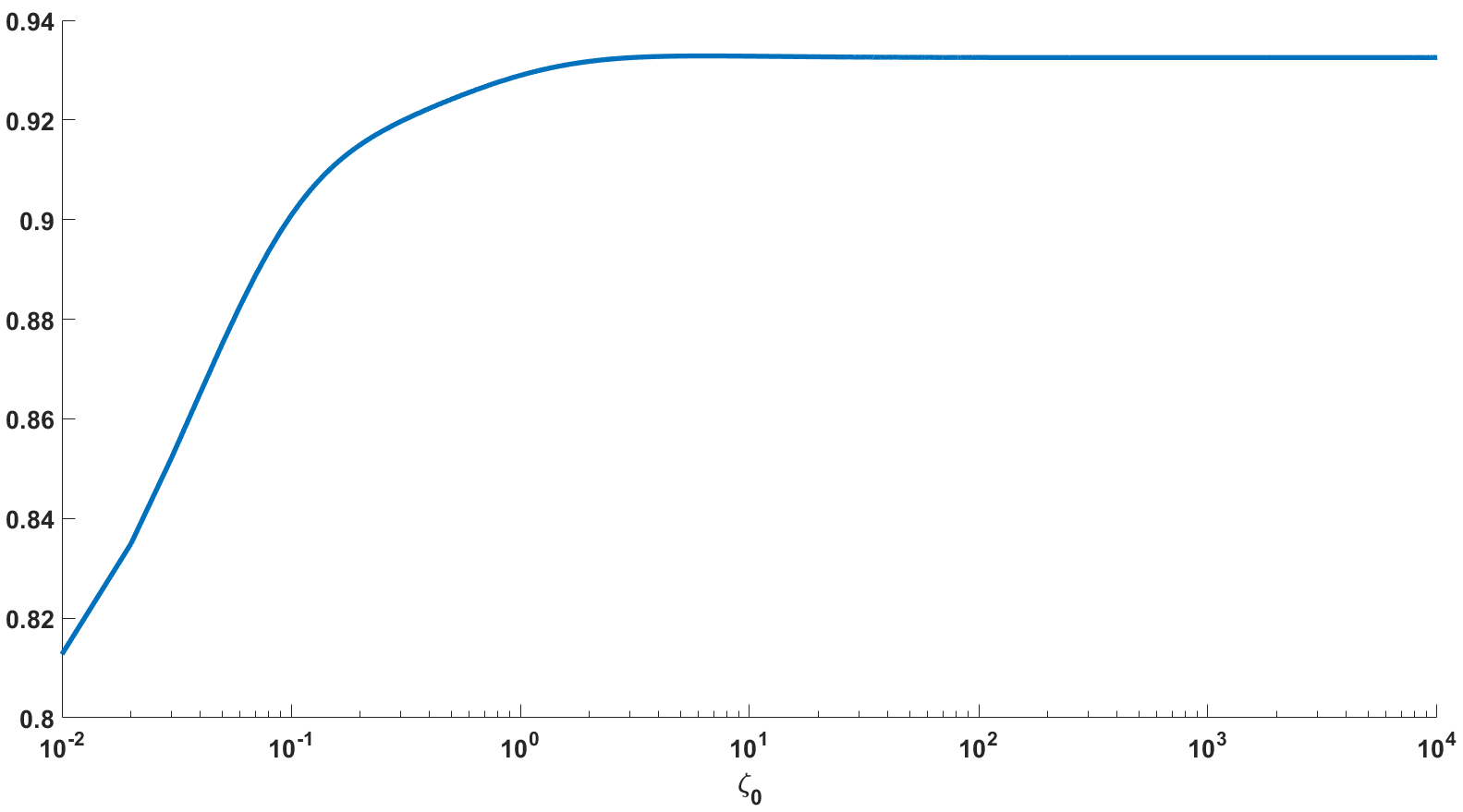}}
	\caption{Integral of $\mathcal{F}_{2}$ over $\left[ 0 , 1 \right]$. $\zeta_{0}$ axis in logarithmic scale. \label{F2_integral}}
\end{figure}

Therefore, we keep up to the quadratic term in the expansion of $W$,
\begin{equation}
	W \left( t \right) \approx i \left\langle u \right\rangle t - \frac{1}{2} \sigma^{2} t^{2} \;\;\; \Rightarrow \;\;\; e^{W \left( t \right)} \approx e^{i \left\langle u \right\rangle t - \frac{1}{2} \sigma^{2} t^{2}}.
\end{equation}
\noindent This approximation leads to 
\begin{equation}
	U_{PV}^{(2)} \left( \zeta_{0} \right) = \frac{2}{1 - \left\langle u \right\rangle } \, \left( \frac{1 - \left\langle u \right\rangle}{\sqrt{2} \sigma} \right) \mathcal{D} \left( \frac{ 1 - \left\langle u \right\rangle}{\sqrt{2} \sigma} \right),
\end{equation}
\noindent where $\mathcal{D}$ is the Dawson function, defined as 
\begin{equation}
	\mathcal{D} (x) \doteq e^{- x^{2}} \int_{0}^{x} ds \, e^{s^{2}}.
	\label{dawson}
\end{equation}
\noindent The mean values $\left\langle u \right\rangle$ and $\left\langle u^{2} \right\rangle$ are 
\begin{equation}
	\left\langle u \right\rangle = \frac{1}{2 \mathcal{K} \left( 0 \right)} \int_{0}^{\infty} dp \int_{\theta_{0}}^{\pi / 2} d \theta \, \frac{g \left( p , \theta \right)}{\text{cosh}^{2} \left( p / 2 \right)} \, \text{tanh} \left[ \zeta_{0} p^{2} \left( 1 - 3 \text{cos}^{2} \theta \right) / 2 \right]
\end{equation}
and
\begin{gather}
	\left\langle u^{2} \right\rangle = \frac{1}{2 \mathcal{K} \left( 0 \right)} \int_{0}^{\infty} dp \int_{\theta_{0}}^{\pi / 2} d \theta \, \frac{g \left( p , \theta \right)}{\text{cosh}^{2} \left( p / 2 \right)} \notag \\
	\times \text{tanh} \left( p / 2 \right) \,  \text{tanh}^{2} \left[ \zeta_{0} p^{2} \left( 1 - 3 \text{cos}^{2} \theta \right) / 2 \right].
\end{gather}
We finally arrive at the final result for the $I_{1}$-type integrals,
\begin{gather}
	I_{1} \left[ g \right] \doteq \int_{0}^{\infty} dp \int_{0}^{\theta_{0}} d \theta \, g(p,\theta) \, f \left( p , \theta \right) \notag \\
	+ \int_{0}^{\infty} dp \, \frac{G \left( p , \theta_{0} \right)}{e^{p} + 1} + \mathcal{K} \left( 0 \right) \left[ U_{PV}^{(2)} \left( \zeta_{0} \right) - 1 \right].
\end{gather}
We now turn to $I_{2}$-type integrals. The same way as before, we keep up to the quadratic term in the expansion of $W$,
\begin{gather}
	U_{PV}^{(2)} \left( \zeta_{0} \right) = \frac{4}{\left( 1 - \left\langle u \right\rangle \right)^{2}} \, \left( \frac{1 - \left\langle u \right\rangle}{\sqrt{2} \sigma} \right)^{3} \mathcal{D} \left( \frac{ 1 - \left\langle u \right\rangle}{\sqrt{2} \sigma} \right) - \frac{1}{\sigma^{2}},
\end{gather}
\noindent where $\mathcal{D}$ is the Dawson function defined in Eq. \eqref{dawson} and the relevant mean values are
\begin{gather}
	\left\langle u \right\rangle = \frac{1}{4 \, \mathcal{K} \left( \zeta_{0} \right)} \int_{0}^{\infty} dp \int_{\theta_{0}}^{\pi / 2} d \theta \, g \left( p , \theta \right) \notag \\
	\times \frac{ \text{tanh} \left( p/2 \right) \text{tanh} \left[ \zeta_{0} p^{2} \left( 1 - 3 \text{cos}^{2} \theta \right) / 2 \right] }{\text{cosh}^{2} \left( p/2 \right) \text{cosh}^{2} \left[ \zeta_{0} p^{2} \left( 1 - 3 \text{cos}^{2} \theta \right) / 2 \right]}
\end{gather}
\noindent and
\begin{gather}
	\left\langle u^{2} \right\rangle = \frac{1}{4 \, \mathcal{K} \left( \zeta_{0} \right)} \int_{0}^{\infty} dp \int_{\theta_{0}}^{\pi / 2} d \theta \, g \left( p , \theta \right) \notag \\
	\times \frac{ \text{tanh}^{2} \left( p/2 \right) \text{tanh}^{2} \left[ \zeta_{0} p^{2} \left( 1 - 3 \text{cos}^{2} \theta \right) / 2 \right] }{\text{cosh}^{2} \left( p/2 \right) \text{cosh}^{2} \left[ \zeta_{0} p^{2} \left( 1 - 3 \text{cos}^{2} \theta \right) / 2 \right]}.
\end{gather}
We finally arrive at the final result for the $I_{2}$-type integrals,
\begin{equation}
	I_{2} \left[ g \right] \doteq \int_{0}^{\infty} dp \int_{0}^{\theta_{0}} d \theta \, g(p,\theta) \, f \left( p , \theta \right) \, \left[ 1 - f \left( p , \theta \right) \right] + \mathcal{K} \left( \zeta_{0} \right) \, U_{PV}^{(2)} \left( \zeta_{0} \right).
\end{equation}
To conclude, let us discuss whether approximating $F_{1}$ and $F_{2}$ by a Gaussian distribution is quantitatively correct.
Let us call $\mathcal{F}_{i}$, $i = 1,2$ the Gaussian approximation of $F_{i}$, given by
\begin{equation}
	\mathcal{F}_{i} \left( u \right) = \frac{1}{\sqrt{2 \pi} \sigma_{i}} e^{- \frac{\left( u - \langle u \rangle_{i} \right)^{2}}{2 \sigma^{2}_{i}}} \;\;\; u \in \mathbf{R},
\end{equation}
\noindent where $\langle u \rangle_{i}$ and $\sigma^{2}_{i} = \langle u^{2} \rangle_{i} - \langle u \rangle^{2}_{i}$ are computed with $F_{i}$.
In the limit $\zeta_{0} \rightarrow 0$, both $F_{1}$ and $F_{2}$ converge to the Dirac Delta distribution,
\begin{equation}
	F_{i} \left( u \right) \xrightarrow[\zeta_{0} \rightarrow 0] \;\;\; \delta \left[ u \right]
\end{equation}
\noindent and, since $\langle u \rangle_{i} \rightarrow 0$ and $\sigma^{2}_{i} \rightarrow 0$, the Gaussian approximations converge to the same limit,
\begin{equation}
	\lim_{\zeta_{0} \to 0} \mathcal{F}_{i} \left( u \right) = \lim_{\sigma_{i} \to 0} \frac{1}{\sqrt{2 \pi} \sigma_{i}} e^{- \frac{ u^{2}}{2 \sigma^{2}_{i}}} = \delta \left[ u \right].
\end{equation}
\noindent Therefore, the limit $\zeta_{0} \rightarrow 0$ is exact.
Arbitrary values of $\zeta_{0}$ require numerical methods to analyze. Figure (\ref{F1}) shows a comparison between $F_{1}$ and its Gaussian approximation $\mathcal{F}_{1}$ as functions of $\zeta_{0}$, using a function $g$
\begin{equation}
	g \left( p, \theta \right) = p^{3} \, \text{sin}^{3} \theta.
\end{equation}
\noindent It shows that for small $\zeta_{0}$ values both $F_{1}$ and $\mathcal{F}_{1}$ tend to a Dirac Delta distribution centered at $u=0$ but for large $\zeta_{0}$ values $F_{1}$ diverges logarithmically at $u=1$ while $\mathcal{F}_{1}$ tends to a Gaussian distribution with constant mean and variance. Although approximating $F_{1}$ by a Gaussian allows the support of the new distribution to be different than the original interval $\left[ 0 , 1 \right]$, figure (\ref{F1_integral}), which is the integral of $\mathcal{F}_{1}$ over $\left[ 0 , 1 \right]$, shows the area under the curve is mostly (at least $75 \%$ of it) located in that interval. Similarly, figure (\ref{F2}) shows a comparison between $F_{2}$ and its Gaussian approximation $\mathcal{F}_{2}$ as functions of $\zeta_{0}$, using a function $g$
\begin{equation}
g \left( p, \theta \right) = p^{4} \, \text{sin} \theta \, \text{cos} \theta.
\end{equation}
\noindent For small $\zeta_{0}$ values both $F_{2}$ and $\mathcal{F}_{2}$ tend to a Dirac Delta distribution centered at $u=0$ but, unlike $F_{1}$, $F_{2}$ is finite for large $\zeta_{0}$ values. Figure (\ref{F2_integral}) shows the area under the curve of $\mathcal{F}_{2}$ in $\left[ 0 , 1 \right]$ is $80 \%$ for small $\zeta_{0}$ values and reaches a constant value of $93 \%$ for large $\zeta_{0}$ values.
We can conclude that the replacement of $F_{1}$ and $F_{2}$ by their corresponding Gaussian approximation effectively cuts off the integrals in a neighborhood of $u=1$ without the need to include an explicit cut-off, which would add a new dimensionfull parameter to the theory. Moreover, the procedure yields a quantitatively accurate approximation for small $\zeta_0$ (it is exact at $\zeta_0=0$). While there is a loss of accuracy for large values of $\zeta_0$, it must be observed that also the energy-momentum tensor is less sensitive to the exact value of $\zeta_0$ in that range, as we will show in next section.

\section{Pressure Anisotropy in DTTs and in the Grad Approximation}

One interesting way to visualize the relationship of $\zeta^{\mu\nu}$ to the energy-momentum tensor in the fully nonlinear DTT is by considering the pressure anisotropy.

We consider an axisymmetric configuration where, in the rest frame, $T^{2} \zeta_0^{ij}=\mathrm{diag}\left( \zeta_0,\zeta_0,-2\zeta_0\right) $ with $\zeta_0\ge 0$. We define the pressure anisotropy as

\be 
\delta_p=\frac{T_{zz}}{\frac12\left( T_{xx}+T_{yy}\right) }=\frac{1+{3\Pi_{zz}}/{\rho}}{1-{3\Pi_{zz}}/{2\rho} }
\te
For comparison, under Grad approximation $\Pi^{\mu\nu}$ is given by Eq. (\ref{Gradpi}). Assuming for $\xi^{\mu\nu}$ the same form as $\zeta^{\mu\nu}$, we get

\be 
\delta_p\approx\frac{1-\frac 85\xi_0}{1+\frac 45\xi_0}
\te 
It becomes negative for $\xi_0>5/8$ and remains negative thereafter, approaching $\delta_p\to -2$ as $\xi_0\to\infty$.

By contrast, DTTs have a built-in lower limit for the pressure anisotropy, because the deformed Fermi-Dirac distribution Eq. (\ref{FDD}) always has a finite dispersion in $p_z$. For example, and leaving out renormalization issues for the moment, when $\zeta_0\to\infty$ Eq. (\ref{FDD}) becomes $f=\Theta\left( 1-3\cos^ 2\theta\right) $, where $\Theta$ denotes the step function. Factoring out and cancelling a divergent radial integral this leads to $\delta_p=1/4$.

\begin{figure}[h]
	\centerline{\includegraphics[width=17cm]{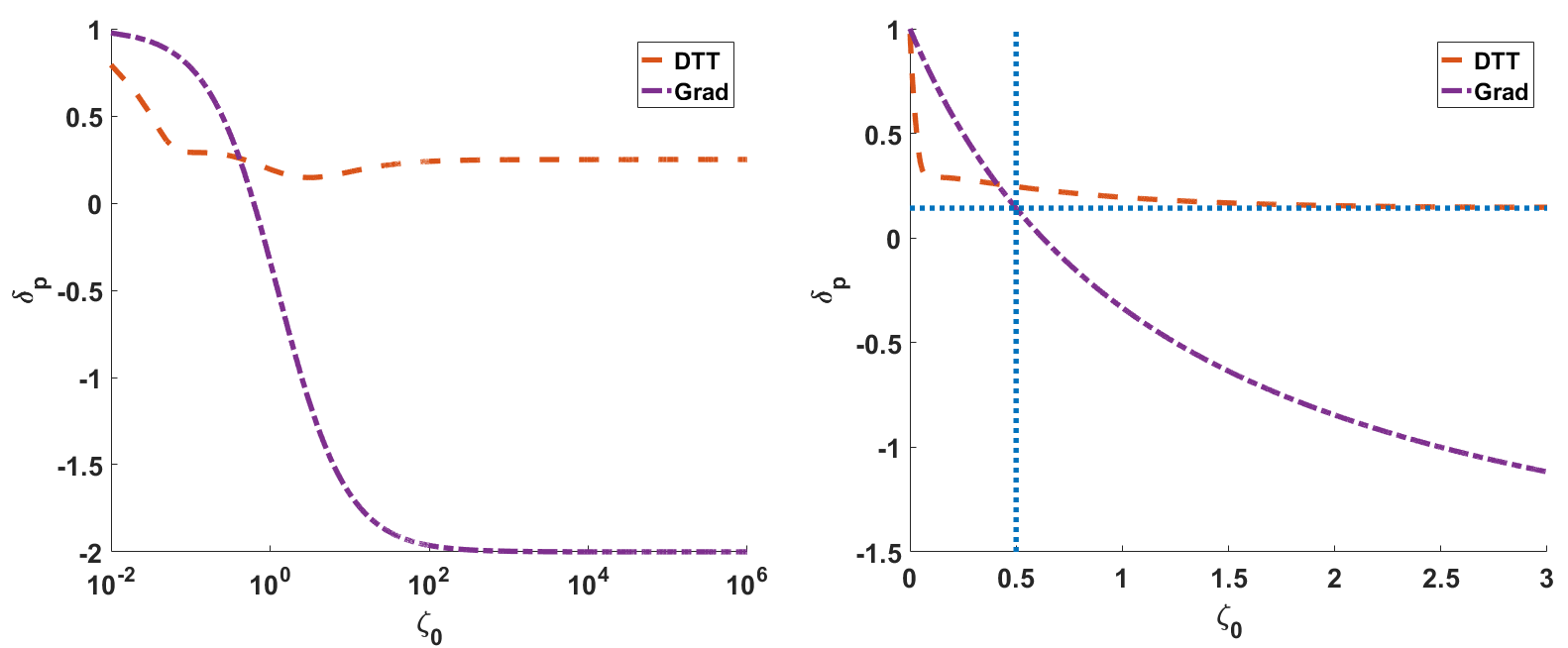}}
	\caption{Left: pressure anisotropy asymptotic behavior. $\zeta_{0}$ axis in logarithmic scale. Right: zoom at the region of interest. Vertical line at $\zeta_{0} = 1/2$ and horizontal line at $\delta_{p} = 1/7$. DTT as a dashed line and Grad's approximation as a dot-dash-dotted line. \label{anisotropy}}
\end{figure}

Of course, a correct evaluation of $\delta_p$ requires that $T^{\mu\nu}$ is computed by carrying out a proper renormalization procedure. When the integrals are regularized by the Gaussian approximation we have presented above, it is seen that $1/4$ is indeed the asymptotic value of $\delta_p$ as $\zeta_0\to\infty$, but that higher anisotropy is possible at finite values. A numerical evaluation shows that $\delta_p\ge 0.1466\approx 1/7$, which in the Grad approximation corresponds to $\xi_0=1/2$, as can be seen in figure (\ref{anisotropy}).

Now let us study the changes in the slope of the DTT anisotropy shown in figure (\ref{anisotropy}) left. If $\delta_{p} = p_{z} / p_{x}$ then the derivative with respect to $\zeta_{0}$ (denoted by primes from now on) is

\be
\delta_{p}^{\prime} = \frac{p_{z}^{\prime}}{p_{x}} \left( 1 - \delta_{p} \frac{p_{x}^{\prime}}{p_{z}^{\prime}} \right),
\te

\noindent where $p_{z}$ and $p_{x}$ are given by

\be
p_{z} = \frac{2}{\left( 2 \pi \right)^{2}} \int_{0}^{\infty} dp \int_{0}^{1} dx \, p^{3} x^{2} \, f
\te

\be
p_{x} = \frac{1}{\left( 2 \pi \right)^{2}} \int_{0}^{\infty} dp \int_{0}^{1} dx \, p^{3} \left( 1 - x^{2} \right) \, f,
\te

\noindent its derivatives $p_{z}^{\prime}$ and $p_{x}^{\prime}$ by

\be
p_{z}^{\prime} = \frac{2}{\left( 2 \pi \right)^{2}} \int_{0}^{\infty} dp \int_{0}^{1} dx \, p^{3} x^{2} \left( 1 - 3x^{2} \right) \, f (1 - f)
\label{pzprime}
\te

\be
p_{x}^{\prime} = \frac{1}{\left( 2 \pi \right)^{2}} \int_{0}^{\infty} dp \int_{0}^{1} dx \, p^{3} \left( 1 - x^{2} \right) \left( 1 - 3x^{2} \right) \, f (1 - f)
\label{pxprime}
\te

\noindent and the distribution function is as in eq. (\ref{FDD}), with $T=1$ for simplicity,

\be
f \left( p , x \right) = \frac{1}{ e^{p - \zeta_{0} p^{2} \left( 1 - 3x^{2} \right)} + 1}.
\te

\noindent In the limit $\zeta_{0} \longrightarrow 0$, $\delta_{p} \longrightarrow 1$ and

\be
\frac{p_{x}^{\prime}}{p_{z}^{\prime}} \longrightarrow \frac{1}{2} \int_{0}^{1} dx \,  \left( 1 - x^{2} \right) \left( 1 - 3x^{2} \right) \, \bigg/ \, \int_{0}^{1} dx \,  x^{2} \left( 1 - 3x^{2} \right) = - \frac{1}{2}.
\te

\noindent Since $p_{z}^{\prime} < 0$ it follows that for sufficiently small $\zeta_{0}$, $\delta_{p}^{\prime} < 0$. Conversely, in the limit $\zeta_{0} \longrightarrow \infty$, $\delta_{p} \longrightarrow 1/4$ and $f (1 - f)$ goes to zero except when $x^{2} \approx 1 /3$. Therefore, for big enough $\zeta_{0}$ values, the integrand in Eqs. (\ref{pzprime}) and (\ref{pxprime}) is concentrated in a neighborhood of $x^{2} = 1 /3$. On the one hand we have

\be
\frac{p_{x}^{\prime}}{p_{z}^{\prime}} \approx \left. \frac{1 - x^{2}}{2 x^{2}} \right|_{x^{2} = 1/3} = 1
\te

\noindent and on the other $p_{z}^{\prime} > 0$ because $f \longrightarrow \Theta\left( 1-3 x^{2} \right) $, so $\delta_{p}^{\prime} > 0$. We have proved that $\delta_{p}$ approaches its asymptotic value from below, in accordance with figure (\ref{anisotropy}) left, which thus captures the general behavior of $\delta_{p}$. In particular, the anisotropy parameter reaches a minimum at some value of $\zeta_0$, regardless of the regularization and renormalization procedure.

This result also shows that the theory can not describe a configuration with anisotropy parameter less than $1/7$; since such configurations would be very extreme, we do not believe this is a significant drawback. At the same time, these are the cases where one would not expect the distribution could be accurately described by a few of its moments. 

For this reason, it seems more important to us that, while an approximation such as Grad's is prone to unphysical behavior in extreme ranges of parameters, the DTT has built in safety measures against such behavior; in this case, no matter how large $\zeta_0$ could become along the evolution, pressures will never become negative. Moreover, the pressure anisotropy is mostly insensitive to the value of $\zeta_0$ when it becomes large, underlying that inaccuracies in the approximations made not necessarily propagate to the physical predictions of the theory.

\section{Perturbative Stability of Homogeneous Configurations}
After outlining the procedure to obtain a well defined DTT out of kinetic theory, we are going to test the resulting theory by considering a problem we can solve both in kinetic theory and in the DTT, and also in a hydrodynamic formalism derived from Grad's approximation. Concretely, we shall discuss whether non equilibrium, homogeneous, anisotropic (but axisymmetric) configurations are perturbatively stable. We shall show this is the case in kinetic theory, meaning that a non-homogeneous perturbation of such a background always decays in time (observe that the background itself is not a solution of the Boltzmann equation). Then we shall obtain a similar result in the DTT by considering the dynamics of linear perturbations to the $u^{\mu}$ and $\zeta_{\mu \nu}$ degrees of freedom. Finally, we shall show that the dynamics of the variables $u^{\mu}$ and $\xi_{\mu \nu}$ from Grad's approximation is unstable if the background is anisotropic enough, even before the lowest pressure actually becomes negative.
We are going to assume an homogeneous temperature $T$ and energy density $\rho$. Following Romatschke and Strickland \cite{RomatschkeStrickland} (see also \cite{Florkowski,Tinti}), $\zeta_{\mu \nu}$ and $\xi_{\mu\nu}$ will be chosen symmetric, transverse and traceless and we will use a coordinate system such that in the unperturbed rest frame, in which $u_0^{\mu} = \delta^{\mu 0}$, the background part of $\zeta_{\mu \nu}$ can be written as $T^2\zeta^{\mu \nu}_{0} = \text{diag} \left( 0 , \zeta_{0} , \zeta_{0} , - 2 \zeta_{0} \right)$, with $\zeta_{0} \geq 0$, and similarly for $\xi_{\mu\nu}$. Since we are perturbing an homogeneous background the normal modes shall be plane waves $e^{st + ikz}$. Our goal will be to find the dispersion relation $s = s \left( k \right)$ by the three formalisms and compare them. An instability appears if for any $k$, $\mathrm{Re}(s)>0$.
\subsection{Kinetic Theory}
We want to solve the kinetic equation for the 1pdf $f$ with an Anderson-Witting collision term Eq. (\ref{kinetic_equation}) 
\cite{Sch06,TI10}. We shall investigate linearized fluctuations around an homogeneous background. To do so, we shall look for the solution of the Boltzmann equation with an initial condition given by a 1pdf of the DTT type Eq. (\ref{FDD}), where moreover the parameters may be decomposed into an homogeneous plus a small, position dependence perturbation. We shall assume this dependence is of the form $e^{ikz}$, in the unperturbed fluid rest frame. At late times, when all transients have decayed, the solution will correspond to a normal mode of the  Boltzmann equation. 

We are going to assume a solution of the form
\begin{equation}
	f = f_{0} \left[ 1 + \left( 1 - f_{0} \right) \delta f e^{ikz}\right],
	\label{kinetic_solution_f}
\end{equation}
\noindent where $f_{0}$ is the background,
\begin{equation}
	f_{0} \left[ p^{\mu} , u_{0}^{\mu} , \zeta_{0}^{\mu \nu} , T \right] = \frac{1}{e^{- \frac{1}{T} {u_{0}}_{\mu} p^{\mu} - \frac{1}{T^{2}} {\zeta_{0}}_{\mu \nu} p^{\mu} p^{\nu}} + 1},
\end{equation}
and $\delta f$ the perturbation. Likewise, $f_{eq}$ is given by
\begin{equation}
	f_{eq} = {f_{eq}}_{0} \left[ 1 + \left( 1 - {f_{eq}}_{0} \right) \delta f_{eq} e^{ikz}\right],
	\label{kinetic_solution_feq}
\end{equation}
where ${f_{eq}}_{0}$ is
\begin{equation}
	{f_{eq}}_{0} \left[ p^{\mu} , u_{0}^{\mu} , T_{eq} \right] = \frac{1}{e^{- \frac{1}{T_{eq}} {u_{0}}_{\mu} p^{\mu}} + 1}.
\end{equation}
If we write the perturbations in the parameters at $t=0$ as
\begin{equation}
	u^{\mu} = u_{0}^{\mu} + v^{\mu}e^{ikz}
	\label{kinetic_u}
\end{equation}
\noindent and
\begin{equation}
	T^{2} \zeta^{\mu \nu} = \zeta_{0}^{\mu \nu} + z^{\mu \nu}e^{ikz},
	\label{kinetic_z}
\end{equation}
\noindent then the perturbation in the initial condition is
\begin{equation}
	\delta f\left(0\right) = \frac{1}{T} \, v_{\mu}(0) p^{\mu} \, + \frac{1}{T^{2}} \, z_{\mu \nu}(0) p^{\mu} p^{\nu} \,
	\label{kinetic_f}
\end{equation}
For the perturbation in $f_{eq}$ we may write
\begin{equation}
	\delta f_{eq} = \frac{1}{T_{eq}} \, v_{\mu}(t) p^{\mu} 
	\label{kinetic_feq}
\end{equation}
\noindent where $T$ and $T_{eq}$ are related by Eq. \eqref{relationT}. Replacing Eqs. \eqref{kinetic_solution_f} and \eqref{kinetic_solution_feq} in Eq. \eqref{kinetic_equation} we arrive at the solution
\begin{gather}
	\delta f (t) = \left( \frac{1}{T} \, v_{\mu} p^{\mu} + \frac{1}{T^{2}} \, z_{\mu \nu} p^{\mu} p^{\nu} \right) e^{- \sigma (p) t} \notag \\
	+ \frac{1}{\tau T_{eq}} F \left( p \right) \, p^{\mu} \int_{0}^{t} dt^{\prime} \, v_{\mu}(t^{\prime}) \, e^{- \sigma \left( t - t^{\prime} \right)},
\end{gather}
\noindent where $\sigma$ and $F$ are given by
\begin{equation}
	\sigma (p) = \frac{1}{\tau} + ik \frac{p_{3}}{p}
\end{equation}
\noindent and
\begin{equation}
	F \left( p \right) = \frac{{f_{eq}}_{0} \left( 1 - {f_{eq}}_{0} \right)}{f_{0} \left( 1 - f_{0} \right)} + \frac{T_{eq}}{p} \left[ \frac{f_{0} - {f_{eq}}_{0}}{f_{0} \left( 1 - f_{0} \right)} \right].
\end{equation}
We will assume the only nonzero perturbations are $v_{1}$ and $z_{13}$. In order to preserve the transversality condition $v_{\mu} \zeta^{\mu \nu} = 0$, $z_{01}$ should be nonzero as well and equal to $z_{01} = - v^{1} \zeta_{0}$. Under this assumptions the solution is
\begin{gather}
	\delta f (t) = \left[ \frac{1}{T} v_{1} + \frac{2}{T^{2}} \left( z_{1 3} p_{3} - \zeta_{0} v_{1} p \right) \right] p_{1} \, e^{- \sigma (p) t} \notag \\
	+ \frac{1}{\tau T_{eq}} F \left( p \right) \, p_{1} \int_{0}^{t} dt^{\prime} \, v_{1}(t^{\prime}) \, e^{- \sigma \left( t - t^{\prime} \right)}.
	\label{kinetic_solution}
\end{gather}
In order to find the dispersion relation we are going to study the long-time behavior of the velocity perturbation. First, recall that the hydrodynamic velocity can be defined as the timelike eigenvector of the energy-momentum tensor, $u_{\mu} T^{\mu \nu} = - \rho u^{\nu}$. For the perturbations we are considering, there are no first order corrections to the energy density. Therefore, up to first order we have
\begin{equation}
	\left( {u_{0}}_{\mu} + v_{\mu} e^{ikz}\right) \left( T^{\mu \nu}_{0} + \delta T^{\mu \nu} \right) = - \rho_{0} \left( u^{\nu}_{0} + v^{\nu} e^{ikz}\right) \;\;\; \Rightarrow \;\;\; v_{1} = \frac{\delta T^{1 0}e^{-ikz}}{\rho_{0} + p_{0}},
	\label{kinetic_velocity}
\end{equation}
\noindent where $p_{0}$ is the equilibrium pressure and $\delta T^{\mu \nu}$ the non-equilibrium part of the energy-momentum tensor. Replacing Eq. \eqref{kinetic_solution} in Eq. \eqref{kinetic_velocity},
\begin{gather}
	v_{1} (t) = \frac{1}{\rho_{0} + p_{0}} \int \frac{d^{3}p}{\left(2 \pi \right)^{3}} \, \left( p_{1} \right)^{2} \left\{ \left[ \frac{1}{T} v_{1} + \frac{2}{T^{2}} \left( z_{1 3} p_{3} - \zeta_{0} v_{1} p \right) \right] \, e^{- \sigma (p) t} \right. \notag \\ %
	\left. + \frac{1}{\tau T_{eq}} F \left( p \right) \, \int_{0}^{t} dt^{\prime} \, v_{1}(t^{\prime}) \, e^{- \sigma \left( t - t^{\prime} \right)} \right\} \, f_{0} \left[ 1 - f_{0} \right].
\end{gather}
\noindent In the limit $t \rightarrow \infty$ we obtain the asymptotic behavior $v_{1} (t)\propto e^{st}$, provided
\begin{equation}
	\text{Re} \left( H \left[ s , k , \zeta_{0} \right] \right) = 1,
	\label{kinetic_implicit}
\end{equation}
\noindent where 
\begin{equation}
	H \left[ s , k , \zeta_{0} \right] = \frac{1}{\rho_{0} + p_{0}} \int \frac{d^{3}p}{\left(2 \pi \right)^{3}} \, \left( p_{1} \right)^{2} \frac{F \left( p \right)}{\tau T_{eq} \left( \sigma + s \right)} f_{0} \left[ 1 - f_{0} \right].
\end{equation}
We introduce a new parameter $\gamma$ defined by
\begin{equation}
	\gamma = \frac{\tau k}{1 + \tau s}.
\end{equation}
\noindent With the aid of $\gamma$, the implicit relation Eq. \eqref{kinetic_implicit} can be written in parametric form as
\begin{equation}
	\begin{dcases}
		\tau s = \mathcal{H} \left[ \gamma , \zeta_{0} \right] - 1 \\ %
		\tau k = \gamma \, \mathcal{H} \left[ \gamma , \zeta_{0} \right]
	\end{dcases}
\end{equation}
\noindent where 
\begin{equation}
	\mathcal{H} \left[ \gamma , \zeta_{0} \right] = \frac{1}{\rho_{0} + p_{0}} \frac{1}{\left( 2 \pi \right)^{2}} \int_{0}^{\infty} dp \int_{0}^{\pi/2} d \theta \, \frac{p^{3} \, \text{sin}^{3} \theta}{1 + \gamma^{2} \text{cos}^{2} \theta} \, \left[ f_{0} + 3 {f_{eq}}_{0} \right].
	\label{kinetic_H}
\end{equation}
\noindent If $\gamma = 0$, then the first term of Eq. \eqref{kinetic_H} is equal to $p_{0} / \left( \rho_{0} + p_{0} \right)$, while the second is just $\rho_{eq} / \left( \rho_{0} + p_{0} \right)$. Due to the fact that $\rho = \rho_{eq}$, $\mathcal{H} \left[ \gamma = 0 , \zeta_{0} \right] = 1$. Therefore, for $\gamma = 0$,
\begin{equation}
	\begin{dcases}
		\tau s = 0 \\ %
		\tau k = 0
	\end{dcases}
\end{equation}
\noindent Since $\mathcal{H}$ is a decreasing function of $\gamma$, we have $\tau s \leq 0$ and, as a consequence, $s$ is  a decreasing function of $k$ with initial value $s(k=0) = 0$. Finally, $s(k) \leq 0$ for any $\gamma$ and we arrive at the conclusion that the linear theory is stable.
\subsection{DTT}
According to the DTT, the dynamics of the gas are governed by the set of equations (\ref{EMT}), (\ref{conservation}), (\ref{NEC}) and (\ref{dtteq}), where $f$ and $f_{eq}$ are defined as in Eqs. (\ref{FDD}) and (\ref{FDeq}), respectively, not just at $t=0$, but at all times. This means that we have the decompositions Eqs. (\ref{kinetic_u}), (\ref{kinetic_z}), (\ref{kinetic_f}) and (\ref{kinetic_feq}) not just at $t=0$, but at all times. As before, we assume the only non zero perturbations are $v_{1}$, $z_{13}$ and $z_{01} = - v^{1} \zeta_{0}$. Linearization leads to the system of equations
\begin{equation}
	\begin{dcases}
		T \left( s C^{\mu \nu} + ik D^{\mu \nu} \right) v_{\nu} + \left( s E^{\mu \nu \rho} + ik F^{\mu \nu \beta} \right) z_{\nu \rho} = 0 \\
		T \left[ \left( s + \frac{1}{\tau} \right) E^{\mu \nu \rho} + ik F^{\mu \nu \rho} \right] v_{\rho} + \left[ \left( s + \frac{1}{\tau} \right) G^{\mu \nu \rho \sigma} + ik H^{\mu \nu \rho \sigma} \right] z_{\rho \sigma}  = 0
	\end{dcases}
\end{equation}
\noindent where the coefficients $C$, $D$, $E$, $F$, $G$ and $H$ are defined by the integrals
\begin{gather}
	C^{\mu \nu} =   \int \, Dp \, p^{\mu} p^{\nu} p^{0} \, f_{0} \left( 1 - f_{0} \right) \;\;\; D^{\mu \nu} = \int \, Dp \, p^{\mu} p^{\nu} p^{3} \, f_{0} \left( 1 - f_{0} \right) \notag \\ %
	E^{\mu \nu \rho} = \int \, Dp \, p^{\mu} p^{\nu} p^{\rho} p^{0} \, f_{0} \left( 1 - f_{0} \right) \;\;\; F^{\mu \nu \rho} = \int \, Dp \, p^{\mu} p^{\nu} p^{\rho} p^{3} \, f_{0} \left( 1 - f_{0} \right) \notag \\ %
	G^{\mu \nu \rho \sigma} = \int \, Dp \, p^{\mu} p^{\nu} p^{\rho} p^{\sigma} p^{0} \, f_{0} \left( 1 - f_{0} \right) \;\;\; H^{\mu \nu \rho \sigma} = \int \, Dp \, p^{\mu} p^{\nu} p^{\rho} p^{\sigma} p^{3} \, f_{0} \left( 1 - f_{0} \right) \notag
\end{gather}
\noindent By keeping the only relevant equations for the $v_{1}$ and $z_{13}$ perturbations (and $z_{01}$ because of the transversality condition), we arrive at the two-by-two linear system
\begin{equation}
	\begin{dcases}
		s \left( T C^{1 1} - 2 \zeta_{0} E^{0 1 1} \right) v_{1} + 2 ik F^{1 1 3} z_{1 3} = 0 \\ %
		ik \left( T F^{1 1 3} - 2 \zeta_{0} H^{0 1 1 3} \right) v_{1} + 2 \left( s + \frac{1}{\tau} \right) H^{0 1 1 3} z_{1 3} = 0
	\end{dcases}
\end{equation}
\noindent Since $T$ is the only dimensionful parameter, there is no loss of generality in setting $T=1$. The dispersion relation is given by the secular equation whose solutions are
\begin{equation}
	\tau s = - \frac{1}{2} \left[ 1 \pm \sqrt{ 1 - 4 \left( \tau k \right)^{2} \Omega_{DTT} \left( \zeta_{0} \right) } \right],
	\label{DTT_dispersion}
\end{equation}
\noindent where  $\Omega_{DTT}$ is defined by
\begin{equation}
	\Omega_{DTT} \left( \zeta_{0} \right) = \frac{F^{113}}{H^{0113}} \left( \frac{F^{113} - 2 \zeta_{0} H^{0113}}{C^{11} - 2 \zeta_{0} E^{011}} \right).
	\label{omegadtt}
\end{equation}
\noindent It is evident that the only interesting case is
\begin{equation}
	\tau s = - \frac{1}{2} \left[ 1 - \sqrt{ 1 - 4 \left( \tau k \right)^{2} \Omega_{DTT} \left( \zeta_{0} \right) } \right],
\end{equation}
\noindent If $\Omega_{DTT} \left( \zeta_{0} \right) < 0$ for some $\zeta_{0}$, then an instability arise. To see if such $\zeta_{0}$ exist, we use the following properties of $f_{0}$,
\begin{equation}
	\frac{\partial f_{0}}{\partial p} = - \left[ 1 - 2 \zeta_{0} \, p \, \left( 1 - 3 \text{cos}^{2} \theta \right) \right] f_{0} \left( 1 - f_{0} \right)
\end{equation}
\noindent and
\begin{equation}
	\frac{\partial f_{0}}{\partial \theta} = 6 \, \zeta_{0} \, p^{2} \text{cos} \theta \, \text{sin} \theta \, f_{0} \left( 1 - f_{0} \right),
\end{equation}
\noindent to write
\begin{equation}
	\left( 1 - 2 \, \zeta_{0} \, p \right) \, f_{0} \left( 1 - f_{0} \right) = - \frac{\partial f_{0}}{\partial p} - \frac{1}{p} \frac{\text{cos} \theta}{\text{sin} \theta} \frac{\partial f_{0}}{\partial \theta}.
\end{equation}
\noindent Using this identity we can integrate by parts to obtain
\begin{equation}
	F^{113} - 2 \zeta_{0} H^{0113} = \frac{2}{\left( 2 \pi \right)^{2}} \int_{0}^{\infty} dp \int_{0}^{\pi/2} d \theta \, p^{4} \, \text{sin} \theta \, \text{cos}^{2} \theta \, f_{0} > 0
\end{equation}
\noindent and
\begin{equation}
	C^{11} - 2 \zeta_{0} E^{011} = \frac{1}{\left( 2 \pi \right)^{2}} \int_{0}^{\infty} dp \int_{0}^{\pi/2} d \theta \, p^{3} \, \text{sin} \theta \, \left( 2 + \text{sin}^{2} \theta \right) \, f_{0} > 0.
\end{equation}
\noindent Since $F^{113} / H^{0113} > 0$, we conclude that $\Omega_{DTT} > 0$ and, just like kinetic theory, perturbations in DTT are stable.
\subsection{Grad's Approximation}
In order to use Grad's approximation first we need to find Grad's probability density function. As discussed in the Introduction, it takes the form Eqs. (\ref{Grad_f}) and (\ref{Grad_Z}). As before, we assume that $T$ is unperturbed, Eq. (\ref{kinetic_u}) for the velocity, and
\be
	\xi^{\mu \nu} = \xi_{0}^{\mu \nu} + x^{\mu \nu}e^{ikz},
	\label{grad_x}
\te
where $\xi_{0}^{\mu \nu}=\mathrm{diag}\left(0,\xi_0,\xi_0,-2\xi_0\right)$, $\xi_0\ge 0$. As before, the only nonzero components of the perturbed variables are $v^1$, $x^{13}=x^{31}$ and $x^{01}=\xi_0v_1$.

The system's dynamics are governed by Eqs.  (\ref{EMT}), (\ref{conservation}), (\ref{NEC}) and (\ref{dtteq}). Replacing Grad's probability density function Eq. \eqref{Grad_f} in the previous equations and solving them for $v_{1}$ and $x_{13}$ up to first order we obtain
\begin{equation}
	\begin{dcases}
		s \left( 1 + \frac{1}{5} \xi_{0} \right) v_{1} + \frac{1}{5} i k \, x_{1 3} = 0 \\ %
		ik \left( 1 - 2 \xi_{0} \right) v_{1} +  \left(s + \frac{1}{\tau} \right) x_{1 3} = 0
	\end{dcases}
\end{equation}
\noindent The dispersion relation is given by the secular equation whose solutions are
\begin{equation}
	\tau s = - \frac{1}{2} \left[ 1 \pm \sqrt{ 1 - 4 \left( \tau k \right)^{2} \Omega_{Grad} \left( \xi_{0} \right)  } \right]
	\label{Grad_dispersion}
\end{equation}
\noindent where $\Omega_{Grad}$ is defined by
\begin{equation}
	\Omega_{Grad} \left( \xi_{0} \right) = \frac{1 - 2 \xi_{0}}{ 5 + \xi_{0}}.
	\label{omegagrad}
\end{equation}
\noindent If we chose the negative sign in Eq. \eqref{Grad_dispersion} and $\xi_{0} > 1/2$, then perturbations show an exponential growth with coefficient
\begin{equation}
	\tau s =  \frac{1}{2} \left[  \sqrt{ 1 + 4 \left( \tau k \right)^{2} \lvert \Omega_{Grad} \left( \xi_{0} \right) \rvert  }-1 \right]
\end{equation}
\noindent There is not a minimum (or maximum) $k$ value for instabilities. They occur at every value of $k$ as long as $\xi_{0} > 1/2$. As we have shown in the previous section, there is a range $1/2<\xi_0\le 5/8$ where all three pressures are positive, but nevertheless this spurious instability appears. For larger $\xi_0$ the lowest pressure becomes negative, bringing the breakdown of the theory to the fore.
\subsection{Quantitative Comparison}
\begin{figure}[h]
	\centerline{\includegraphics[width=17cm]{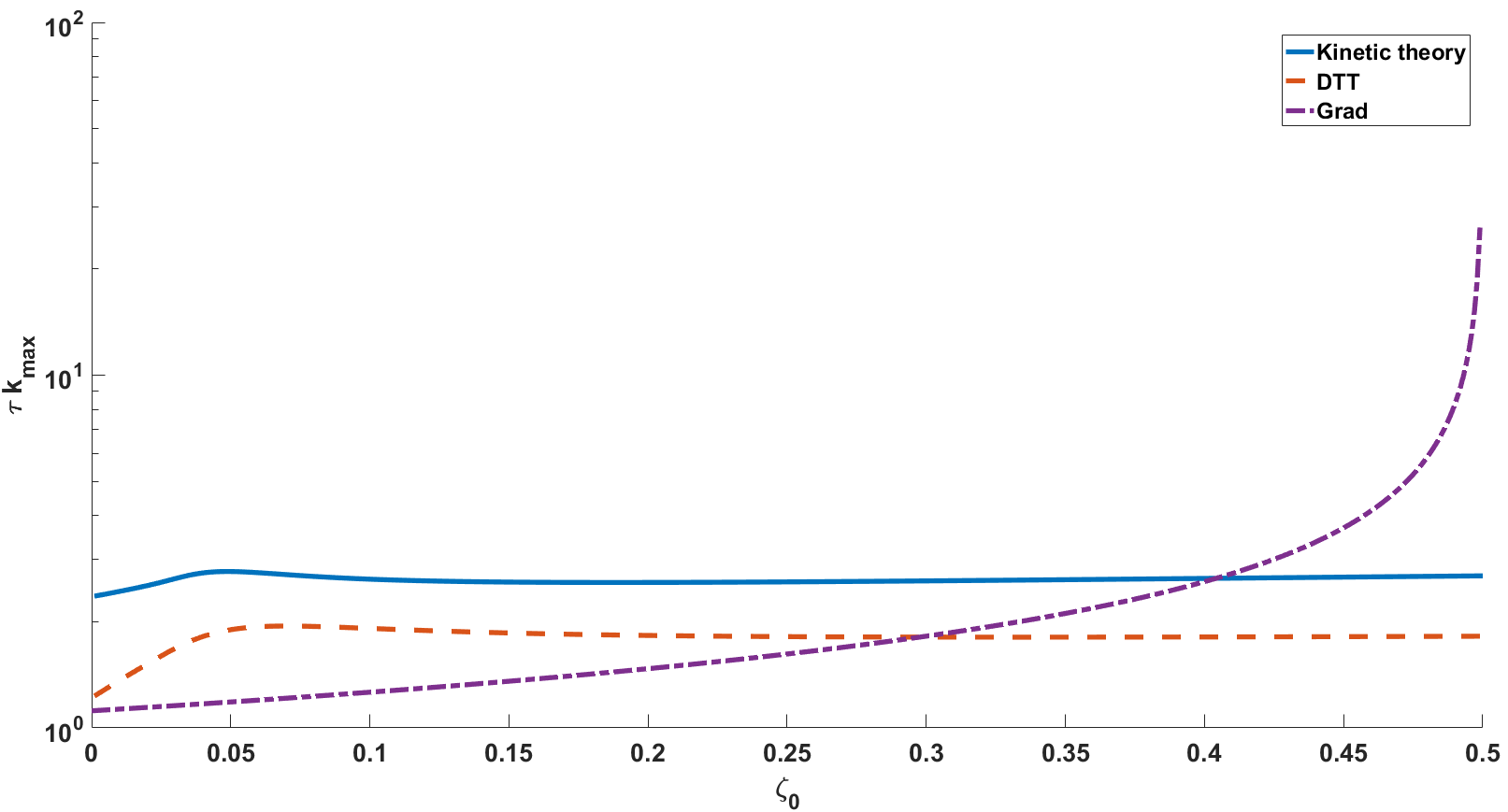}}
	\caption{$\tau k_{max}$ as a function of $\zeta_{0}=\xi_0 $. It shows kinetic theory as a full line, DTT as a dashed line and Grad's approximation as a dot-dash-dotted line. $\tau k_{max}$ axis in logarithmic scale. \label{kmax}}
\end{figure}

\begin{figure}[h]
	\centerline{\includegraphics[width=17cm]{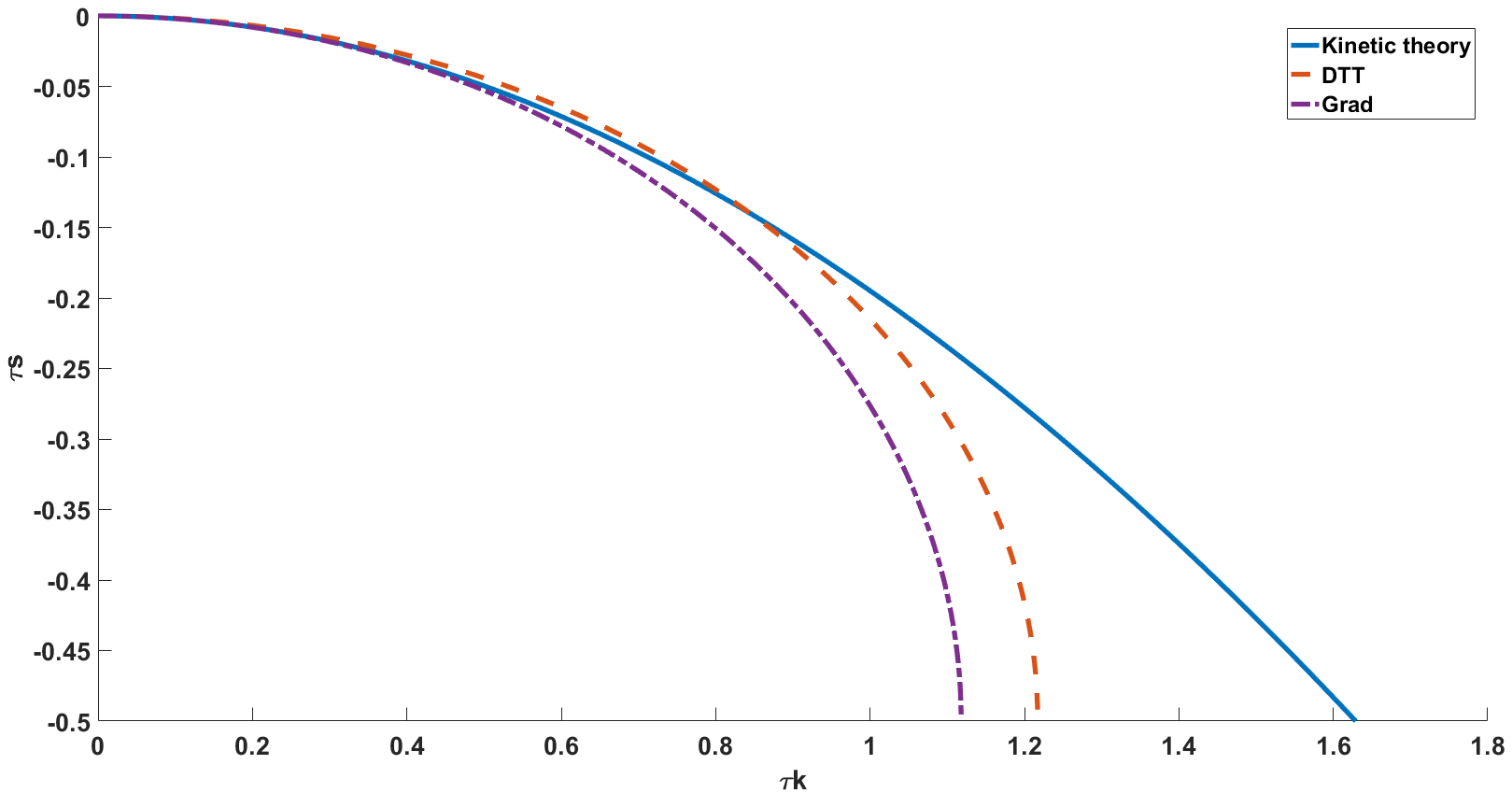}}
	\caption{Dispersion relation for $\zeta_{0}=\xi_0  = 0$. It shows kinetic theory as a full line, DTT as a dashed line and Grad's approximation as a dot-dash-dotted line. \label{dispersion_0}}
\end{figure}

\begin{figure}[h]
	\centerline{\includegraphics[width=17cm]{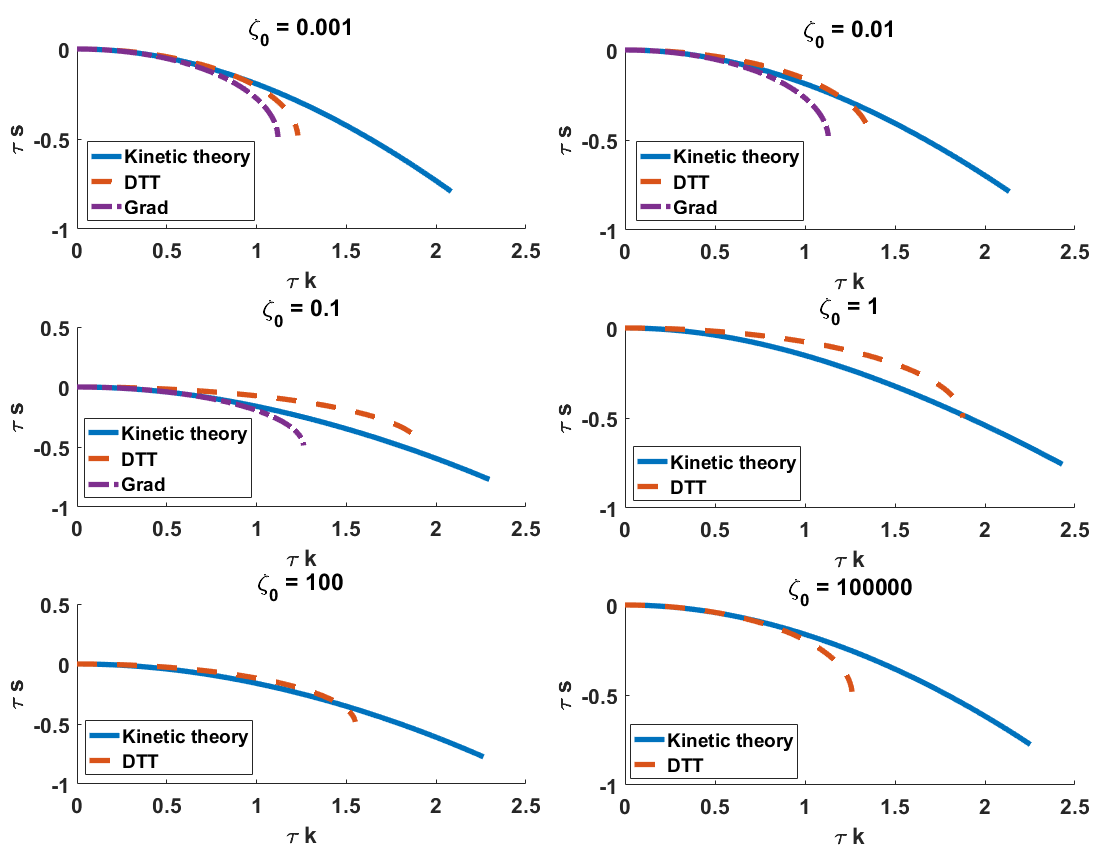}}
	\caption{Dispersion relation for different $\zeta_{0}=\xi_0 $ values. It shows kinetic theory as a full line, DTT as a dashed line and Grad's approximation as a dot-dash-dotted line. \label{dispersion}}
\end{figure}

\begin{figure}[h]
	\centerline{\includegraphics[width=17cm]{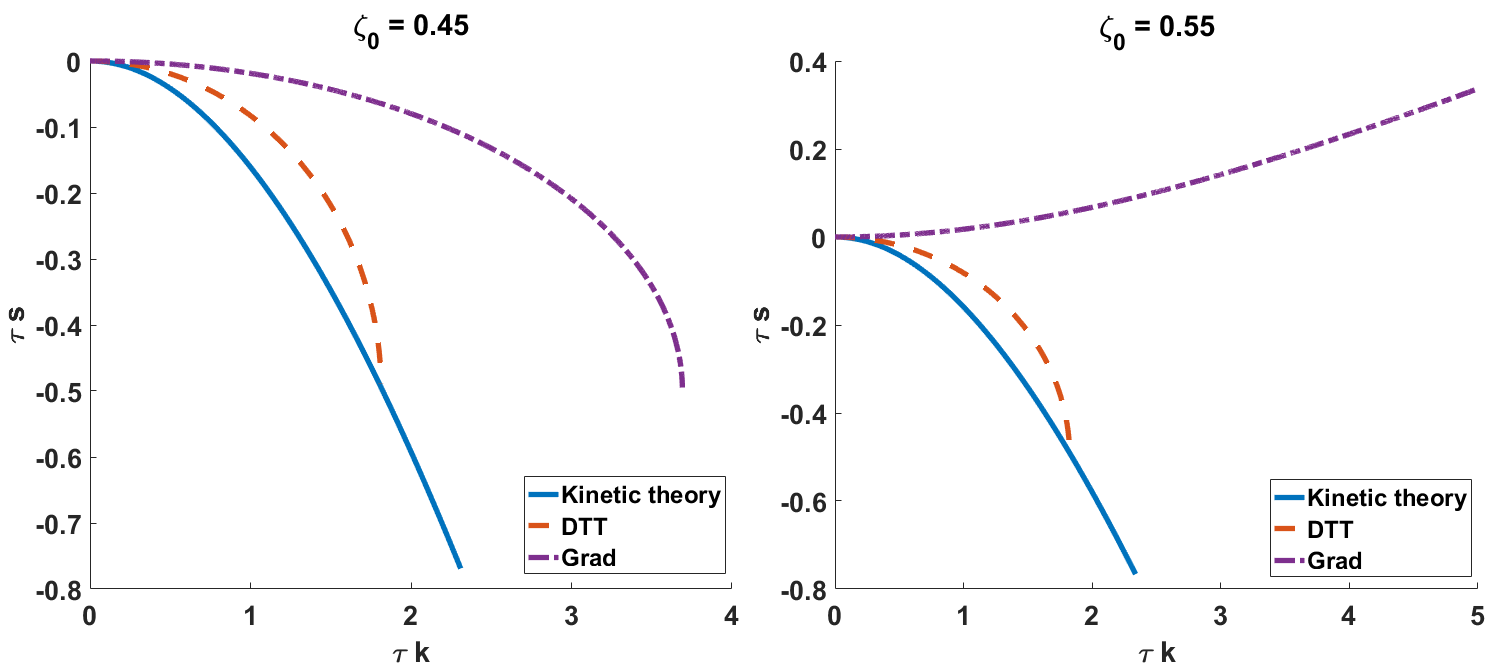}}
	\caption{Dispersion relation in a neighborhood of Grad's instability. It shows kinetic theory as a full line, DTT as a dashed line and Grad's approximation (where $\xi_0=\zeta_0$) as a dot-dash-dotted line. \label{grad}}
\end{figure}

The main difference between the three methods utilized before is the presence or absence of instabilities. While kinetic theory and DTT show no signs of them, Grad's approximation has no stable solutions for $\xi_{0} > 1/2$. This is a drawback for Grad since it shows its applicability is fairly limited.
All three theories predict an $s_{max} < 0$ with its corresponding $k_{max}$ (also in Grad's approximation $\xi_{0}$ must be less than $1/2$) such that if $k > k_{max}$ then the dispersion relation can be written as
\begin{equation}
	s(k) = s_{max} + i h(k) \;\;\; h(k) \in \mathbf{R}.
\end{equation}
\noindent This means we have propagation in the form of damped waves. In kinetic theory this set of $s_{max}$ and $k_{max}$ are given by
\begin{equation}
	\tau s \xrightarrow[\gamma \rightarrow \infty] \;\;\; \tau s_{max} =  - 1
\end{equation}
\noindent and
\begin{gather}
	\tau k \xrightarrow[\gamma \rightarrow \infty] \;\;\; \tau k_{max} = \frac{1}{8 \pi} \left[ \frac{1}{\rho_{0} + p_{0}} \right. \notag \\
	\times \left. \int_{0}^{\infty} dp \, p^{3} \left( \frac{1}{e^{\frac{1}{T} p - \frac{1}{T^{2}} \zeta_{0} p^{2}} + 1} -  \frac{3}{e^{\frac{1}{T_{eq}}p} + 1}  \right) \right].
\end{gather}
\noindent Both DTT and Grad have the same $s_{max}$,
\begin{equation}
	\tau s_{\max} = - \frac{1}{2},
\end{equation}
\noindent but they have a different $k_{max}$. In DTT it's given by
\begin{equation}
	\tau k_{max} = \frac{1}{2 \sqrt{\Omega_{DTT} \left( \zeta_{0} \right)}}
\end{equation}
\noindent while in Grad its value is
\begin{equation}
	\tau k_{max} = \frac{1}{2 \sqrt{\Omega_{Grad} \left( \xi_{0} \right)}} \;\;\; \left( \xi_{0} < 1/2\right) .
\end{equation}
Figure (\ref{kmax}) shows a comparison between the three different values of $k_{max}$ as a function of the anisotropy $\zeta_{0},\xi_0$. It can be seen that DTT shows a qualitatively similar behavior to kinetic theory, better than Grad's.
The isotropic case, that is $\zeta_{0}=\xi_0 = 0$, can be solved analytically. Up to second order in $k$ we have, in kinetic theory
\begin{equation}
	\tau s \approx - 0.5 \left( \tau k \right)^{2},
\end{equation}
\noindent in DTT,
\begin{equation}
	\tau s \approx - 0.71 \left( \tau k \right)^{2}
\end{equation}
\noindent and in Grad's approximation,
\begin{equation}
	\tau s \approx - 0.2 \left( \tau k \right)^{2}.
\end{equation}
Figure (\ref{dispersion_0}) shows the full dispersion relations for $\zeta_{0}=\xi_0 = 0$, showing that even up to $\tau k \approx 0.9$ DTT looks very similar to kinetic theory.
Arbitrary values of $\zeta_{0}$ and $\xi_0 $ require numerical methods to solve, always utilizing the regularization procedure defined in previous sections. Figure (\ref{dispersion}) shows the dispersion relations for all three theories as functions of $\zeta_{0}=\xi_0 $. For both small and big values of $\zeta_{0}$, DTT is a good approximation to kinetic theory. Figure (\ref{grad}) shows the dispersion relation for $\zeta_{0}=\xi_0  = 0.45$ and $\zeta_{0}=\xi_0  = 0.55$, that is, before and after Grad's instability.
\section{Final Remarks}

Formulating a fully nonlinear hydrodynamics of dissipative relativistic fluids is a daunting challenge that we must nevertheless confront if we wish to make sense of the very early stages of relativistic heavy ion collisions \cite{Str14b,JeHe15,Rom16} and also of the cosmic evolution in the period that goes, roughly, from reheating after inflation to the electroweak and QCD transitions \cite{BTW06,AHKK15,BVS06,CK10,CK14}. We believe this paper contributes to the ongoing effort to meet this challenge in two main ways. On the one hand, it delineates the boundary of applicability of a representative ``second order'' theory. These theories were introduced to solve the instability problems of the so-called ``first order'' theories \cite{HisLin83,HiscockLindblom,HisLin88}; nevertheless, as we have shown, they display spurious instabilities of their own. On the other hand, we show a definite way whereby a fully nonlinear hydrodynamics may be derived from kinetic theory in a systematic way.

Of course this is not the only strategy that is being tried out \cite{MNDR14}. The best known is simply to go to higher orders within the Chapman-Enskog or Grad approaches, as in the so-called Burnett's equations \cite{Burnett}; the second order approximation is discussed in \cite{Jais13b,Jais14,Jais15,Jais15b}. These models very soon become extremely complex, which may become an issue if we consider that the kind of problems we have discussed in this paper are already an oversimplification of the problems we really want to study, and which include gauge and possibly the gravitational field as well.

A promising strategy is to duplicate the Grad approach but taking as zeroth order an already nonequilibrium state, as in the so-called anisotropic hydrodynamics \cite{Str14b,BHS14,BNR16}. To the best of our knowledge this approach has been tried so far only in highly symmetric configurations\cite{TRFS15,FRST16,FMR15}, so it is unclear what hurdles it could encounter in realistic scenarios.

In our view, the DTT framework we are advocating has two distinctive advantages. First, it enforces energy-momentum conservation and the Second Law in a rigorous way, contrary to ``second order'' theories in which it is enforced only to second order. In typical ``second order'' theories (see however \cite{Jaiswal} and \cite{MartinezStrickland}), the entropy production, as computed from the entropy flux and the hydrodynamic equations, is nonnegative only if terms of order higher than second are neglected \cite{Cristian}. In a DTT, the same procedure yields an strictly nonnegative expression to all orders in deviation from ideal behavior. Second, it can describe situations far from equilibrium without the addition of other degrees of freedom than those already present in $T^{\mu\nu}$; this puts a limit on how complex the theory may become, although of course it will never be as simple and compelling as the hydrodynamics of ideal fluids. Moreover, having a fully consistent theory to begin with gives one a solid framework whereby one can discuss simplifications in a systematic way (by contrast, observe that the Eckart expansion is known not to be convergent \cite{ChapmanEnskogDivergence,Resurgence}). It could well be that the main value of the theory we have developed in this paper is that it exists, rather than its actual applications.

In last analysis, to be able to compare several alternative formalisms will be a definite asset for the community as we enter in this largely uncharted territory. 

\section*{Appendix A: Stability in conformal hydrodynamics}
We wish to provide a general template for the discussion of stability against incompressible perturbations in conformal hydrodynamical theories, where the fundamental equations are energy-momentum conservation and a new conservation law of type eq. (\ref{DDT_dynamics2}) for some totally symmetric (and traceless on any pair of indexes) tensor $A^{\mu\nu\rho}$. Following Israel and Stewart \cite{IsrSt79}, we call an hydrodynamical theory one where the fundamental degrees of freedom are in one-to-one correspondence with the components of the energy-momentum tensor, at least in a neighborhood of the equilibrium states. Therefore, since we are restricting ourselves to conformal theories, the fundamental degrees of freedom can be chosen as a single dimensionful scalar $T$ (which becomes the temperature in equilibrium states), the Landau-Lifshitz velocity $u^{\mu}$ and a dimensionless, symmetric, traceless and transverse tensor $Z^{\mu\nu}$. Since $Z^{\mu\nu}$ is transverse, we cannot build new tensors by contracting it with $u^{\mu}$; the only other linearly independent transverse traceless tensor in the theory is $\hat{Z}^{\mu\nu}=Z^{2\mu\nu}-\left(1/3\right)\mathrm{tr}Z^2\;\Delta^{\mu\nu}$. It follows that we have the decomposition

\bea
T^{\mu\nu}&=&T^4\left\{A_T\left[u^{\mu}u^{\nu}+\frac13\Delta^{\mu\nu}\right]+BZ^{\mu\nu}+B'\hat{Z}^{\mu\nu}\right\}\nn
A^{\mu\nu\rho}&=&T^5\left\{A_A\left[u^{\mu}u^{\nu}u^{\rho}+
\frac13\left(\Delta^{\mu\nu}u^{\rho}+\Delta^{\nu\rho}u^{\mu}+\Delta^{\rho\mu}u^{\nu}\right) \right]\right.\nn &+&\left. C\left(Z^{\mu\nu}u^{\rho}+Z^{\nu\rho}u^{\mu}+Z^{\rho\mu}u^{\nu}\right)+C'\left(\hat{Z}^{\mu\nu}u^{\rho}+\hat{Z}^{\nu\rho}u^{\mu}+\hat{Z}^{\rho\mu}u^{\nu}\right)\right\}\nn
I^{\mu\nu}&=&T^6\left\{DZ^{\mu\nu}+D'\hat{Z}^{\mu\nu}\right\}
\tea
The scalars $A_T,A_A,B,C,D,B',C'$ and $D'$ are functions of $Z^{\mu\nu}$ through invariants such as $\mathrm{tr}Z^2$ and $\mathrm{tr}Z^3$.

We consider linear perturbations to an homogeneous anisotropic background. This means quantity $X$ becomes $X=X_{background}+\delta X\;e^{st+ikz}$. In the background $u^{\mu}=\delta^{0\mu}$ and $Z^{\mu\nu}=\mathrm{diag}\left( 0,Z_0,Z_0,-2Z_0\right) $ ($\mathrm{tr}Z^2=6Z_0^2$). The only perturbed component of the velocity is $\delta u^1=v$. The perturbed components of $Z^{\mu\nu}$ are $\delta Z^{13}=\delta Z^{31}=z$ and $\delta Z^{01}=\delta Z^{10}=Z_0v$, as demanded by transversality. It follows that $\delta \hat{Z}^{13}=\delta \hat{Z}^{31}=-Z_0z$ and $\delta \hat{Z}^{01}=\delta \hat{Z}^{10}=-Z_0^2v$, all other zero. $T$ and all the invariants constructed from $Z^{\mu\nu}$ are unchanged. The relevant equations of motion

\bea 
s\delta T^{01}+ik\delta T^{31}&=&0\nn
s\delta A^{031}+ik\delta A^{331}&=&\delta I^{31}
\tea
become 

\bea 
s\left[ \frac43A_T+\left( B-B'Z_0\right) Z_0\right] v+ik\left( B-B'Z_0\right)z&=&0\nn
s\left( C-C'Z_0\right)z+ik\left[ \frac13A_A-2\left( C-C'Z_0\right)Z_0\right] v&=&T\left( D-D'Z_0\right) z
\tea
leading to the dispersion relation (compare to eqs. (\ref{DTT_dispersion}) and (\ref{Grad_dispersion}))

\be 
\tau s=\frac{-1}2\left\lbrace 1\pm\sqrt{1-4\Omega \left( \tau^2 k^2\right) }\right\rbrace 
\te 
where 

\bea 
\tau &=&\frac{-\left( C-C'Z_0\right)}{T\left( D-D'Z_0\right) }\nn
\Omega &=&\frac{\left( B-B'Z_0\right)\left[ \frac13A_A-2\left( C-C'Z_0\right)Z_0\right] }{\left( C-C'Z_0\right)\left[ \frac43A_T+\left( B-B'Z_0\right) Z_0\right]}
\tea
If $\tau <0$ the theory is always unstable for long wavelenghts, which is clearly unphysical. If $\tau >0$, the theory becomes unstable for short wavelenghts if $\Omega <0$. So stability requires both $\tau >0$ and $\Omega >0$.

However this condition cannot be met if we force $T^{\mu\nu}$ and $A^{\mu\nu\rho}$ to be linear functions of $Z^{\mu\nu}$. This amounts to defining $B'=C'=D'=0$ and $A_T, A_A, B, C$ and $D$ to be constants. Moreover stability at equilibrium implies that $\Omega\left( Z_0=0\right) =B A_A/4C A_T>0$, so $C/A_A$ and $B/A_T$ must have the same sign. Therefore $\Omega $ becomes a rational function which cannot be nonnegative everywhere. This is what happens in the Grad theory, where, after identifying $Z_0=\xi_0$, we obtain $A_T=\sigma_{SB}$, $B/A_T=4/15$ and $C/A_A=1/3$. $\Omega =\Omega_{Grad}$ is given by eq. (\ref{omegagrad}), which is clearly negative for $\xi_0>1/2$.

In the DTT, on the other hand, we have $Z_0=\zeta_0$ and $\Omega =\Omega_{DTT}$ given by eq. (\ref{omegadtt}), which as we have seen is indeed positive for $\zeta_0>0$.

\begin{figure}[h]
	\centerline{\includegraphics[width=17cm]{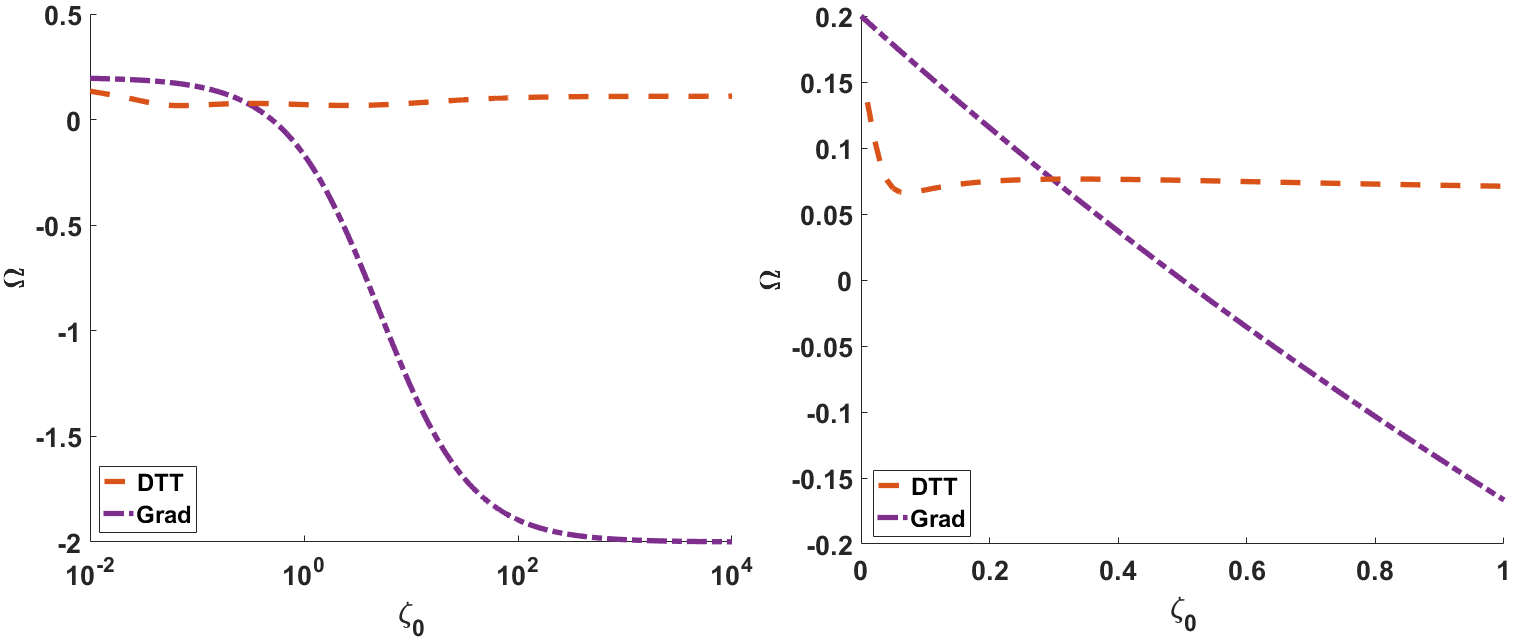}}
	\caption{$\Omega$ for both Grad and DTT. Logarithmic $\zeta_{0}$ axis on the left and zoom on the sign change interval on the right. \label{omega}}
\end{figure}

We show both $\Omega_{DTT}$ and $\Omega_{Grad}$ as functions of $\zeta_0=\xi_0$ in figure (\ref{omega}). Both start from a positive value at $\zeta_0=0$ with a negative slope, but while $\Omega_{Grad}$ is monotonous and eventually reaches the asymptotic value of $-2$, $\Omega_{DTT}$ changes its tendency and remains positive for all $\zeta_0>0$. This underlies that the instability of the equations from the Grad approximation is an artifact of the linearization of $T^{\mu\nu}$ and $A^{\mu\nu\rho}$ with respect to $\xi^{\mu\nu}$. 

\section*{Appendix B: Extension to Maxwell-J\"uttner and Bose-Einstein statistics}

In this appendix we will show that the same regularization and renormalization procedure we have applied to Fermi-Dirac (FD) particles may be used for particles obeying  Maxwell-J\"uttner (MJ) or Bose-Einstein (BE) statistics. Our starting point is the observation that all three statistics may be obtained as particular cases of theories described by the family of one-particle distribution functions 

\begin{equation}
	f_a \left( p , \theta \right) = \frac{1}{ e^{p - \zeta_{0} p^{2} \left( 1 - 3 \text{cos}^{2} \theta \right)} + e^{-a}}.
\end{equation}
where $a=0$ for FD, $a\to\infty$ yields MJ and $a=\pm i\pi$ gives BE. The idea is to obtain the expectation value of some function $g(p,\theta)$ (as in the main text) as a function of $a$ for $a$ real and positive, and then try and extend the result to the MJ and BE cases. We only consider case I above (cfr. eq. (\ref{Regularizacion_I1}))

\begin{equation}
	I_{1a} \left[ g \right] = \int_{0}^{\infty} dp \int_{0}^{\pi / 2} d \theta \, g(p,\theta) \, f_a \left( p , \theta \right)
	\label{Regularizacion_I1a}
\end{equation}

As in the main text, we start by dividing the  $I_{1a}$ integral in $\theta_{0} = \text{cos}^{-1} \left( 1 / \sqrt{3} \right)$,
\begin{gather}
	I_{1a} \left[ g \right] = I_{1a}^{<} \left[ g \right] + I_{1a}^{>} \left[ g \right] \notag \\
	\doteq \int_{0}^{\infty} dp \int_{0}^{\theta_{0}} d \theta \, g(p,\theta) \, f_a \left( p , \theta \right) + \int_{0}^{\infty} dp \int_{\theta_{0}}^{\pi / 2} d \theta \, g(p,\theta) \, f_a \left( p , \theta \right).
\end{gather}
$I_{1a}^{<}$ is well defined in all three cases and will be left as is. To study the $I_{1a}^{>}$ term, define a function $G$ as in the main text (eq. (\ref{gfunction}))
and integrate by parts to obtain
\begin{gather}
	I_{1a}^{>} = \int_{0}^{\infty} dp \, \frac{G \left( p , \theta_{0} \right)}{e^{p} + e^{-a}}  \notag \\ %
	+ \frac{3}{2} \zeta_{0}e^{a} \int_{0}^{\infty} dp \int_{\theta_{0}}^{\pi / 2} d \theta \, p^{2} \, G \left( p , \theta \right) \, \frac{\text{cos} \theta \, \text{sin} \theta}{\text{cosh}^{2} \left[ \left( p+a\right) /2 - \zeta_{0} p^{2} \left( 1 - 3 \text{cos}^{2} \theta \right) / 2 \right]}.
\end{gather}
\noindent it is evident that the first term is finite in all three cases. Let's call the second term $K_{Za}$. By using the sum of arguments relation of the hyperbolic cosine and realizing it could be written as a partial derivative, it is possible to rewrite $K_{Za}$ as
\begin{gather}
	K_{Za} = e^{a}\int_{0}^{\infty} dp \int_{\theta_{0}}^{\pi / 2} d \theta \, \frac{G \left( p , \theta \right)}{\text{sinh}  \left( p+a\right)} \notag \\
	\times \frac{\partial}{\partial \theta} \left\{ \frac{1}{ 1 - \text{tanh} \left( \left( p+a\right)/2 \right) \text{tanh} \left[ \zeta_{0} p^{2} \left( 1 - 3 \text{cos}^{2} \theta \right) / 2 \right] } \right\}.
\end{gather}
\noindent Performing another integration by parts and defining the auxiliary function
\begin{equation}
	\mathcal{K}_a \left( \zeta_{0} \right) \doteq \int_{0}^{\infty} dp \int_{\theta_{0}}^{\pi / 2} d \theta \, \frac{g \left( p , \theta \right)}{\text{sinh} \left( p+a\right)} \,  \frac{1}{ 1 - \text{tanh} \left( \left( p+a\right)/2 \right) \text{tanh} \left[ \zeta_{0} p^{2} \left( 1 - 3 \text{cos}^{2} \theta \right) / 2 \right] },
\end{equation}
\noindent we arrive at the final expression for $K_{Za}$,
\begin{equation}
	K_{Za} = e^{a}\mathcal{K}_a \left( 0 \right) \left[ \frac{\mathcal{K}_a \left( \zeta_{0} \right)}{\mathcal{K}_a \left( 0 \right)} - 1 \right].
	\label{KZa}
\end{equation}
\noindent The key here is to identify the ratio in Eq. \eqref{KZa} as the mean value
\begin{equation}
	U_a \left( \zeta_{0} \right) \doteq \frac{\mathcal{K}_a \left( \zeta_{0} \right)}{\mathcal{K}_a \left( 0 \right)} = \left\langle \frac{1}{1 - u} \right\rangle = \int du \, \frac{F_{1a} \left( u \right)}{1 - u},
\end{equation}
\noindent where $F_{1a}$ is a probability density function defined as
\begin{gather}
	F_{1a} \left( u \right) \doteq \frac{1}{\mathcal{K}_a \left( 0 \right)} \int_{0}^{\infty} dp \int_{\theta_{0}}^{\pi / 2} d \theta \, \frac{g \left( p , \theta \right)}{\text{sinh} \left( p+a \right)} \notag \\ 
	\times \delta \left[ \text{tanh} \left( \left( p+a\right) /2 \right) \text{tanh} \left[ \zeta_{0} p^{2} \left( 1 - 3 \text{cos}^{2} \theta \right) / 2 \right] - u \right]
\end{gather}
We leave open the range of $u$. As in the main text we replace $U_a$ by its Cauchy principal value,
\begin{equation}
	U_a \left( \zeta_{0} \right) \rightarrow 	U_{aPV} \left( \zeta_{0} \right) = \text{Re} \left[ \text{PV} \int du \, \frac{F_{1a} \left( u \right)}{1 - u} \right],
\end{equation}
The idea is to approximate $F_{1a}$ by a Gaussian, for which we need the mean values $\left\langle u \right\rangle_a$ and $\left\langle u^{2} \right\rangle_a$ 
\begin{equation}
	\left\langle u \right\rangle_a = \frac{1}{2 \mathcal{K}_a \left( 0 \right)} \int_{0}^{\infty} dp \int_{\theta_{0}}^{\pi / 2} d \theta \, \frac{g \left( p , \theta \right)}{\text{cosh}^{2} \left( \left( p+a\right)  / 2 \right)} \, \text{tanh} \left[ \zeta_{0} p^{2} \left( 1 - 3 \text{cos}^{2} \theta \right) / 2 \right]
\end{equation}
and
\begin{gather}
	\left\langle u^{2} \right\rangle_a = \frac{1}{2 \mathcal{K}_a \left( 0 \right)} \int_{0}^{\infty} dp \int_{\theta_{0}}^{\pi / 2} d \theta \, \frac{g \left( p , \theta \right)}{\text{cosh}^{2} \left( \left( p+a\right)  / 2 \right)} \notag \\
	\times \text{tanh} \left( \left( p+a\right)  / 2 \right) \,  \text{tanh}^{2} \left[ \zeta_{0} p^{2} \left( 1 - 3 \text{cos}^{2} \theta \right) / 2 \right].
\end{gather}
The limit $a\to\infty$ is not problematic. If $a=\pm i\pi$ we have the identities 

\bea 
\sinh \left( p\pm i\pi\right) &=& -\sinh p\nn
\cosh \left( p\pm i\pi\right) &=& -\cosh p\nn
\sinh \left( \left( p\pm i\pi\right) /2\right) &=& \pm i\cosh p/2\nn
\cosh \left( \left( p\pm i\pi\right) /2\right) &=& \pm i\sinh p/2\nn
\tanh \left( \left( p\pm i\pi\right) /2\right) &=& \left[ \tanh p/2\right] ^{-1}
\tea
Again, the relevant expectation values are well defined.

\section*{Acknowledgments}
Work supported in part by CONICET and University of Buenos Aires. It is a pleasure to thank R. Ferraro, A. Jaiswal, A. Kandus, F. Lombardo, P. Mininni, N. Mir\'on y C. Vega for discussions.

\end{document}